\documentclass[%
reprint,
 amsmath,amssymb,
 aps,
]{revtex4-2}

\usepackage{subfigure}
\usepackage{float}
\usepackage{slashed}
\usepackage{xcolor}
\usepackage{amsmath}
\usepackage{subcaption}
\usepackage[english]{babel}
\usepackage{tikz}
\usepackage{adjustbox}
\usepackage{rotating}
\usepackage{rotfloat}
\usepackage{verbatim}
\usepackage[utf8]{inputenc}
\usepackage{graphicx}
\usepackage{dcolumn}
\usepackage{bm}
\usepackage[super]{nth}
\usepackage{lipsum}

\usepackage{hyperref}
\usepackage{xcolor}
\hypersetup{
  colorlinks   = true, 
  urlcolor     = red, 
  linkcolor    = blue, 
  citecolor   = blue 
}

\usepackage{threeparttable}
\begin{document}

\preprint{APS/123-QED}

\title{Spin identification of the mono-Z$^{\prime}$ resonance in muon-pair production at the ILC with simulated electron-positron collisions at $\sqrt{s}$ = 500 GeV}

\author{S. Elgammal}
 \altaffiliation[sherif.elgammal@bue.edu.eg]{}
\affiliation{Centre for Theoretical Physics, The British University in Egypt, P.O. Box 43, El Sherouk City, Cairo 11837, Egypt.} 


\date{\today}

\begin{abstract}
{In this analysis, we examine the angular distribution of low-mass dimuon pairs produced in simulated electron-positron collisions at the proposed International Linear Collider (ILC), which operates at a center-of-mass energy of 500 GeV and has an integrated luminosity of 4 ab\(^{-1}\).
Our focus is on the cos\(\theta_{\text{CS}}\) variable, which is defined in the Collins-Soper frame. In the standard model, the production of low-mass dimuon pairs is primarily driven by the Drell-Yan process, which exhibits a notable forward-backward asymmetry. However, many scenarios beyond the standard model predict different shapes for the cos\(\theta_{\text{CS}}\) distribution. This angular distribution can be valuable for distinguishing between these models, especially in the event of observing excesses beyond the standard model expectations.
We utilized the mono-Z\(^{\prime}\) model to interpret the simulated data. In the absence of any discoveries of new physics, we establish upper limits at the 95\% confidence level on the masses of various particles in the model, which includes the spin-1 \(Z^{\prime}\) boson and fermionic dark matter.}

\vspace{0.75cm}
\end{abstract}

\maketitle

\section{Introduction}
\label{sec:intro}
One promising approach for detecting physics beyond the standard model (SM) at future electron-positron colliders involves examining changes in the dilepton mass spectrum. These changes might appear as a new peak, which could be predicted by models featuring neutral gauge bosons, such as the Z$^{\prime}$ \cite{heaveyZ}, or theories like Randall-Sundrum \cite{extradim}. Alternatively, the spectrum may show a broader distortion. Such distortions could hint at the presence of Contact Interactions \cite{leptoquark, ContactInteraction} or models like ADD \cite{ADD}. To support these theories, the mass spectrum should reveal an excess or a deficit of events compared to the background prediction, which is predominantly shaped by the Drell-Yan process.

The collaboration of CMS has conducted a detailed study of signatures related to Z$^{\prime}$ and Contact Interaction models \cite{ZprimeandCI}. The ATLAS and CMS collaborations have searched for the massive extra neutral gauge boson Z$^{\prime}$, which is predicted by Grand Unified Theory (GUT) and Supersymmetry. \cite{Extra-Gauge-bosons, gaugeboson1, LR-symmetry11, Super-symmetry12}. However, there is currently no evidence after analyzing the entire Run 2 period of LHC data \cite{ZprimeandCI, zprimeATLAS}. 
The results from the CMS experiment have excluded the existence of Z$^{\prime}$, at a 95\% Confidence Level (CL), for mass values ranging from 0.6 to 5.15 TeV, while the ATLAS experiment has ruled out mass values between 0.6 and 5.1 TeV.

Previous collider experiments, such as LEP-2 \cite{lep}, have established constraints on the masses of Z$^{\prime}$ bosons at a center-of-mass energy of 209 GeV. As a result, LEP-2 has determined lower limits on the mass of a potential Z$^{\prime}$ boson across various models. These limits range from 340 to 1787 GeV, depending on the specific model.

The angular distributions of the leptons are also expected to be affected. Previous studies conducted by the CMS \cite{AfbCMS} and ATLAS \cite{AfbATLAS} collaborations analyzed the angular distributions of Drell-Yan processes involving the production of charged lepton pairs near the Z-boson mass peak, which allows for the measurement of the forward-backward asymmetry denoted as \(A_{FB}\). These studies used the complete dataset from LHC Run 1, which includes an integrated luminosity of 19.7 \(\text{fb}^{-1}\) for CMS and 20.3 \(\text{fb}^{-1}\) for ATLAS, derived from proton-proton collisions at a center-of-mass energy of 8 TeV (\(\sqrt{s}\)). The results indicated that the measurements of \(A_{FB}\) are consistent with the predictions of the standard model.
Furthermore, the forward-backward asymmetry of high-mass dilepton events (with invariant mass \(M_{ll} > 170\) GeV) has been measured using the CMS detector at a center-of-mass energy of \(\sqrt{s} = 13\) TeV, with an integrated luminosity of 138 fb\(^{-1}\). This analysis concluded that no statistically significant deviations from the predictions of the standard model have been observed \cite{AfbCMS13tev}.

Numerous dark matter (DM) searches have been performed by examining data from the CMS and ATLAS experiments during Run 2 \cite{CMS-DM, ATLAS-DM}. These investigations concentrate on the generation of a visible particle, referred to as "X," which experiences a recoil against the significant missing transverse energy produced by dark matter particles. This results in a distinctive signature of \((\text{X} + E^{\text{miss}}_{T})\) recorded in the detector \cite{R38}. The visible particle "X" could be a SM particle, such as W or Z bosons, jets \cite{R35, atlasmonoZ, mono-zCMS, mono-zATLAS}, a photon \cite{photon, photonATLAS}, or even the SM Higgs boson \cite{R36, monoHiggsAtlas1, monoHiggsAtlas2}.

In addition, research was being conducted on dark matter in connection with hard photons (commonly referred to as mono-photons) \cite{monoPhoton-ilc2} at upcoming electron-positron colliders such as the International Linear Collider (ILC) \cite{ilc-report} and the Compact Linear Collider (CLIC) \cite{clic-report}. At the anticipated 500 GeV phase of the ILC, studies suggest that it could effectively detect Weakly Interacting Massive Particles (WIMPs) as dark matter candidates with masses reaching approximately 250 GeV, which is about half of the center-of-mass energy, at a confidence level of 95\%.

Searches for dark matter associated with the Z boson (Mono-Z) at the ILC \cite{monoZ-ilc2} indicate that analyzing dimension-six operators in EFT can reveal insights into leptophilic and/or electrophilic dark matter in the 20 to 400 GeV mass range, as well as the scale of this effective theory.

Earlier experiments with electron-positron colliders, such as LEP-2, have established a lower limit on the mass of dark matter to be approximately 40 GeV. This conclusion is based on the assumption that all soft supersymmetry-breaking scalar masses (such as slepton, squark, and Higgs masses) are universal at a certain Grand Unified Theory (GUT) input scale \cite{DM-lep2}.

Recent collider experiments, such as those conducted by ATLAS and CMS, have imposed strict limits on the interaction of the Z$^{\prime}$ particle with SM leptons, represented as \(\texttt{g}_{l}\).
Based on observations of four-muon final states, the coupling constant \(\texttt{g}_{l}\) has been excluded in the range of 0.004 to 0.3, with variation depending on the mass of the Z$^{\prime}$ boson. \cite{R130,R131}. 

The ATLAS collaboration has conducted a search for dark matter \cite{atlasmonoZprime}   within the framework of the mono-Z$^{\prime}$ model \cite{R1, monoZprime2}, specifically for high Z$^{\prime}$ masses (i.e., \(M_{Z^{\prime}} > 200~\text{GeV}\)). This search focused on the leptonic decay channel of Z$^{\prime}$ at the LHC and has excluded Z$^{\prime}$ masses ranging from 200 to 1000 GeV. Additionally, it has imposed specific constraints on \(\texttt{g}_{l}\).
In the context of the light-vector scenario, the values of \(\texttt{g}_{l}\) are ruled out in the ranges of 0.01 to 0.025 for lower Z$^{\prime}$ masses, and from 0.02 to 0.38 for higher Z$^{\prime}$ masses, specifically those falling between 200 and 1000 GeV.

The LEP-2 \cite{lep} provided constraints on \(\texttt{g}_{l}\). For Z$^{\prime}$ masses exceeding $\sqrt{s} = 209~\text{GeV}$, the limit is given by \(\texttt{g}_{l} \leq 0.044 \times M_{Z^{\prime}}/(\text{200 GeV})\). Conversely, for $M_{Z^{\prime}}$ values below 209~\text{GeV}, the limit remains constant at \(\texttt{g}_{l} \leq 0.044\) \cite{R37}.

If the Z$^{\prime}$ does not interact with quarks, then the HL-LHC and future hadron colliders would not be able to detect such particles. In this case, electron-positron colliders, such as the proposed International Linear Collider (ILC) \cite{ilc1,ilc2,ilc3,ilc4}, will play a pivotal role in deepening our understanding of these phenomena.

The baseline scenario of ILC starts from 250 GeV center-of-mass operation, and 500 GeV is considered as an upgrade.
When the ILC completes its run at $\sqrt{s} = 500$ GeV, it is expected to generate significantly larger data sets totaling 4 ab$^{-1}$. Additionally, the electron beam will be polarized at 80\%, and if the undulator-based positron source concept is implemented, positrons will be provided with a polarization of 30\% as described in the ILC Snowmass Report \cite{ILCSnowmass}.

Linear electron-positron colliders offer controllable energy, reduced QCD background, and adjustable beam polarization.

This analysis investigates light neutral gauge bosons (Z$^{\prime}$) with masses up to 100 GeV, utilizing the light vector (LV) simplified model within a mono-Z$^{\prime}$ framework \cite{R1} at the ILC. We examine simulated electron-positron collisions at the ILC, maintaining a center of mass energy of 500 GeV and an integrated luminosity of 4 ab$^{-1}$. In the following sections, we consistently apply an electron beam polarization of 80\% and a positron beam polarization of 30\%, as outlined in the ILC Snowmass Report \cite{ILCSnowmass}. Our focus lies on dimuon events resulting from Z$^{\prime}$ decay, along with large missing transverse energy that is associated with dark matter.

This paper is structured as follows: Section \ref{section:CS} introduces the Collins-Soper frame and the cos$\theta_{\text{CS}}$ variable for Z boson events. Section \ref{section:model} presents the theoretical framework of the mono-Z$^{\prime}$ portal model. In section \ref{section:MCandDat}, we discuss the simulation techniques for signal and SM background samples. Section \ref{section:AnSelection} covers the selection cuts and analysis strategy. 
Background reduction is discussed in Section \ref{section:BkgReduction}.
Finally, sections \ref{section:Results} and \ref{section:Summary} present the results and summary of the analysis.

\section{The Collins-Soper frame}
\label{section:CS}
When an electron and a positron collide, they can produce a lepton pair ($l^+l^-$). The angle $\theta$ measures the angle between the negative lepton and the incoming electron or positron in the center of the mass frame. The primary process for producing this pair at the tree level in the standard model is the Z boson process ($e^{-} e^{+} \rightarrow Z \rightarrow l^+l^-$). Our analysis focuses on the angular distribution of \( l^+l^- \) pairs in the Collins-Soper frame \cite{CSpaper}, which reduces distortions from the transverse momenta of the colliding particles. 
We use the Collins-Soper frame to analyze the angular distribution of lepton pairs, defining the angle $\theta_{CS}$ as the angle between the negative lepton momentum and the z-axis.

To determine the Collins-Soper frame orientation, we use the sign of the longitudinal boost of the dilepton system. The angle $\text{cos}\theta_{CS}$ can be computed from measurable lab frame quantities, as explained in \cite{AfbCMS}.
\begin{equation}
     \text{cos}\theta_{CS} = \frac{|Q_z|}{Q_z} \frac{2(P_1^+ P_2^- - P_1^- P_2^+)}{\sqrt{Q^2(Q^2 + Q_T^2)}}.     
    \label{costheta:equ}
\end{equation}
The symbols \( Q \), \( Q_T \), and \( Q_z \) represent the four-momentum, transverse momentum, and longitudinal momentum of the dilepton system, respectively. In this context, \( P_1 \) and \( P_2 \) denote the four-momentum of the particles \( l^- \) and \( l^+ \), respectively, while \( E_i \) indicates the energy of each lepton. Additionally, \( P^\pm_i \) is defined as \( \frac{E_i \pm P_{z,i}}{\sqrt{2}} \).

\begin{figure} [h!]
\centering
\resizebox*{6.0cm}{!}{\includegraphics{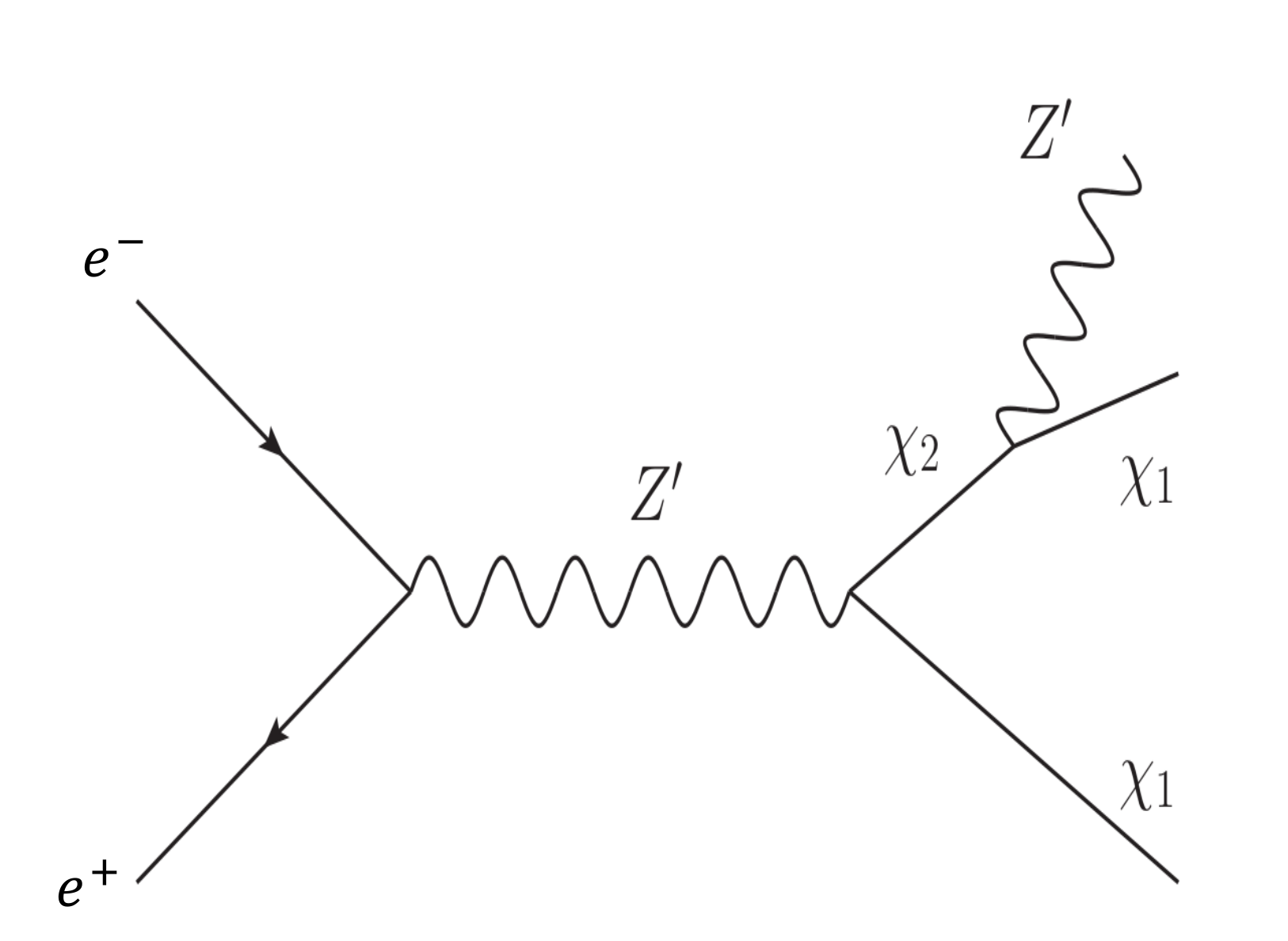}}
\caption{The Feynman diagram for the Light Vector (LV) scenario illustrating the production of a neutral gauge boson (Z$^{\prime}$) alongside a dark matter pair ($\chi_{1}$) \cite{R1}.}
\label{figure:fig1}
\end{figure}

\section{The theoretical model within the context of the mono-Z$^{\prime}$ portal.}
\label{section:model}
The mono-Z$^{\prime}$ model, which has been presented in \cite{R1}, describes dark matter production from electron-positron collisions at the ILC via a new light gauge boson Z$^{\prime}$. It includes two minimal renormalizable scenarios: the first involves dark-Higgsstrahlung from a Z$^{\prime}$ with invisible decay of the dark Higgs. In contrast, the second involves two states, $\chi_{1,2}$, coupling off-diagonally to the Z$^{\prime}$. The search for Z$^{\prime} + E^{miss}_{T}$ is notably more sensitive than direct resonance searches, particularly at lower Z$^{\prime}$ masses. This analysis specifically targets the scenario involving the light vector (LV), as illustrated in Figure \ref{figure:fig1}.

The proposed mechanism for dark matter production involves the annihilation of electron-positron pairs, which is mediated by a light vector boson known as Z$^{\prime}$. This interaction gives rise to two distinct types of dark matter: a lighter form labeled as $(\chi_{1})$ and a heavier variant referred to as $(\chi_{2})$. The heavier dark matter, $(\chi_{2})$, subsequently decays into a Z$^{\prime}$ and the lighter dark matter, $(\chi_{1})$, following the process $\chi_{2} \rightarrow Z^{\prime} + \chi_{1}$.

The interaction term in the Lagrangian for dark fermions and Z$^{\prime}$ is described in \cite{R1}. 
\begin{equation}
    \frac{\texttt{g}_{DM}}{2} Z^{\prime}_{\mu}\large(\bar{\chi_{2}}\gamma^{\mu}\gamma^{5}\chi_{1} + \bar{\chi_{1}}\gamma^{\mu}\gamma^{5}\chi_{2}\large), \nonumber
\end{equation}
where $\texttt{g}_{DM}$ denotes the coupling of Z$^{\prime}$ to dark matter $\chi_{1}$. The coupling of Z$^{\prime}$ to visible leptons is represented by $\texttt{g}_{l}$.
\begin{table} [h!]    
\small
\centering
\begin {tabular} {ll}
\hline
\hline
Scenario & \hspace{3pt} Masses assumptions \\
\hline
\\
    & \hspace{4pt} $M_{\chi_{1}} = 1, 5, ...,200$ GeV \\
Light dark sector & \\
    & \hspace{4pt} $M_{\chi_{2}} = M_{\chi_{1}} + M_{Z'} + 25$ GeV
  \\
 \\
\hline
\hline
\end {tabular}
\vspace{3pt}
\caption{The assumptions regarding the light mass for the dark sector in the LV scenario. \cite{R1}.}
\label{table:tab1}
\end{table}

In the LV scenario, the only permitted decay processes are as follows: ${Z}'\rightarrow\chi_{1}\chi_{2}$, $\chi_{2}\rightarrow {Z'}\chi_{1}$. In this model Z$^{\prime}$ couples only to charged leptons ($e$, $\mu$) and quarks \cite{R1}. 
A Z$^{\prime}$ coupling to electrons is tightly constrained by measurements from LEP \cite{R1, 51}. 
If the Z$^{\prime}$ tends to couple more strongly with muons and quarks, future linear colliders could offer valuable insights that enhance the precision measurements obtained from LEP. For this reason, our analysis focuses on the muonic decay of the Z$^{\prime}$.

The total decay widths of both the ${Z}'$ and ${\chi_{2}}$ can be calculated using the masses of ${Z}'$ and the dark matter, along with the relevant coupling constants.
The free parameters in this scenario include the lightest dark matter mass $M_{\chi_{1}}$, the mass of the second dark matter particle $M_{\chi_{2}}$, the mass of the Z$^{\prime}$ boson $(M_{Z^{\prime}})$, and the couplings of Z$^{\prime}$ to both leptons and dark matter particles, $\texttt{g}_{l}$ and $\texttt{g}_{DM}$, respectively.

Since both the CMS and ATLAS detectors have ruled out Z$^{\prime}$ bosons in the mass range of 0.2 to 5.15 TeV, we focus on the production of lighter neutral gauge bosons (Z$^{\prime}$) below 100 GeV at the ILC.

We consider the LV scenario using the light-dark sector model to provide mass to the dark matter particles ($\chi_{1}$ and $\chi_{2}$), as detailed in Table \ref{table:tab1}. This choice of $M_{\chi_{1}}$ and $M_{\chi_{2}}$ follows a prescription from \cite{R1}, but it is just one possible option.

Due to previous restrictions from experiments like CMS, ATLAS, and LEP-2, the value of $\texttt{g}_{l}$ is approximately 0.003 for $M_{Z^{\prime}}$ between 10 and 100 GeV \cite{R37}. In contrast, $\texttt{g}_{DM}$ is set to 1.0 \cite{R1}, while the masses ($M_{Z^{\prime}}$, $M_{\chi_{1}}$, and $M_{\chi_{2}}$) are varied.

\begin{sidewaystable} []
\centering
\scriptsize
\fontsize{8.pt}{11pt}
\selectfont
\begin{tabular}{|c|c|c|c|c|c|c|c|c|c|c|c|}
\hline
&\multicolumn{11}{|c|}{$M_{Z'}$ (GeV)}\\
\hline
$M_{\chi_{1}}$ (GeV)&10 & 20 & 30 & 40 & 50 & 60 & 70 & 80 & 90 & 100 & 150  \\
\hline
1  & 
    $7.73\times10^{-1}$ & 
    $7.08\times10^{-1}$ & 
    $6.73\times10^{-1}$ & 
    $6.49\times10^{-1}$ & 
    $6.28\times10^{-1}$ & 
    $6.12\times10^{-1}$ & 
    $6.01\times10^{-1}$ & 
    $5.88\times10^{-1}$ &
    $5.78\times10^{-1}$ &
    $5.69\times10^{-1}$ &
    $5.37\times10^{-1}$\\
\hline
5  & 
    $7.63\times10^{-1}$ & 
    $6.87\times10^{-1}$ & 
    $6.52\times10^{-1}$ & 
    $6.29\times10^{-1}$ & 
    $6.09\times10^{-1}$ & 
    $5.95\times10^{-1}$ & 
    $5.82\times10^{-1}$ & 
    $5.70\times10^{-1}$ &
    $5.60\times10^{-1}$ &
    $5.51\times10^{-1}$ &
    $5.17\times10^{-1}$\\
\hline
10  & 
    $7.62\times10^{-1}$ & 
    $6.65\times10^{-1}$ & 
    $6.30\times10^{-1}$ & 
    $6.05\times10^{-1}$ & 
    $5.86\times10^{-1}$ & 
    $5.73\times10^{-1}$ & 
    $5.59\times10^{-1}$ & 
    $5.48\times10^{-1}$ &
    $5.39\times10^{-1}$ &
    $5.29\times10^{-1}$ &
    $4.91\times10^{-1}$\\
    \hline
25  & 
    $6.44\times10^{-1}$ & 
    $6.20\times10^{-1}$ & 
    $5.75\times10^{-1}$ & 
    $5.47\times10^{-1}$ & 
    $5.27\times10^{-1}$ & 
    $5.12\times10^{-1}$ & 
    $4.98\times10^{-1}$ & 
    $4.86\times10^{-1}$ &
    $4.76\times10^{-1}$ &
    $4.66\times10^{-1}$ &
    $4.21\times10^{-1}$\\
    \hline
70  & 
    $5.58\times10^{-1}$ & 
    $5.19\times10^{-1}$ & 
    $4.79\times10^{-1}$ & 
    $4.41\times10^{-1}$ & 
    $4.04\times10^{-1}$ & 
    $3.66\times10^{-1}$ & 
    $3.31\times10^{-1}$ & 
    $2.95\times10^{-1}$ &
    $2.60\times10^{-1}$ &
    $2.42\times10^{-1}$ &
    $2.28\times10^{-1}$ \\
\hline
100 & 
    $6.30\times10^{-1}$ & 
    $3.80\times10^{-1}$ & 
    $3.15\times10^{-1}$ & 
    $2.82\times10^{-1}$ & 
    $2.59\times10^{-1}$ & 
    $2.40\times10^{-1}$ & 
    $2.24\times10^{-1}$ & 
    $2.10\times10^{-1}$ &
    $1.94\times10^{-1}$ &
    $1.81\times10^{-1}$ &
    $1.17\times10^{-1}$\\
\hline
125 & 
    $4.46\times10^{-1}$ & 
    $2.69\times10^{-1}$ & 
    $2.20\times10^{-1}$ & 
    $1.93\times10^{-1}$ & 
    $1.73\times10^{-1}$ & 
    $1.56\times10^{-1}$ & 
    $1.41\times10^{-1}$ & 
    $1.28\times10^{-1}$ &
    $1.17\times10^{-1}$ &
    $1.02\times10^{-1}$ &
    $4.64\times10^{-2}$\\
\hline
130 &  
    $4.08\times10^{-1}$ & 
    $2.47\times10^{-1}$ & 
    $2.01\times10^{-1}$ & 
    $1.76\times10^{-1}$ & 
    $1.56\times10^{-1}$ & 
    $1.40\times10^{-1}$ & 
    $1.26\times10^{-1}$ & 
    $1.12\times10^{-1}$ &
    $1.00\times10^{-1}$ &
    $8.77\times10^{-2}$ &
    $3.53\times10^{-2}$\\

\hline
145 &  
    $2.96\times10^{-1}$ & 
    $1.82\times10^{-1}$ & 
    $1.47\times10^{-1}$ & 
    $1.26\times10^{-1}$ & 
    $1.10\times10^{-1}$ & 
    $9.60\times10^{-2}$ & 
    $8.34\times10^{-2}$ & 
    $7.18\times10^{-2}$ &
    $6.07\times10^{-2}$ &
    $5.05\times10^{-2}$ &
    $1.10\times10^{-2}$\\
    
\hline
170 &  
    $1.35\times10^{-1}$ & 
    $8.95\times10^{-2}$ & 
    $7.07\times10^{-2}$ & 
    $5.77\times10^{-2}$ & 
    $4.67\times10^{-2}$ & 
    $3.72\times10^{-2}$ & 
    $2.86\times10^{-2}$ & 
    $2.12\times10^{-2}$ &
    $1.48\times10^{-2}$ &
    $9.46\times10^{-3}$ &
    $8.39\times10^{-6}$\\
    \hline
178 &  
    $9.61\times10^{-2}$ & 
    $6.60\times10^{-2}$ & 
    $5.11\times10^{-2}$ & 
    $4.02\times10^{-2}$ & 
    $3.12\times10^{-2}$ & 
    $2.34\times10^{-2}$ & 
    $1.66\times10^{-2}$ & 
    $1.10\times10^{-2}$ &
    $6.56\times10^{-3}$ &
    $3.24\times10^{-3}$ &
    $1.46\times10^{-11}$ \\
\hline
200 & 
    $2.48\times10^{-2}$ & 
    $1.84\times10^{-2}$ & 
    $1.28\times10^{-2}$ & 
    $8.08\times10^{-3}$ & 
    $4.50\times10^{-3}$ & 
    $2.02\times10^{-3}$ & 
    $6.27\times10^{-4}$ & 
    $1.15\times10^{-4}$ &
    $8.19\times10^{-6}$ &
    $1.00\times10^{-6}$ &
    $2.63\times10^{-13}$ \\

\hline
204 & 
    $1.90\times10^{-2}$ & 
    $1.30\times10^{-2}$ & 
    $9.74\times10^{-3}$ & 
    $6.54\times10^{-3}$ & 
    $3.62\times10^{-3}$ & 
    $1.50\times10^{-3}$ & 
    $3.99\times10^{-4}$ & 
    $4.75\times10^{-5}$ &
    $1.14\times10^{-7}$ &
    $2.94\times10^{-12}$ &
    $3.03\times10^{-13}$ \\
    
    \hline
\end {tabular}
\caption{The LV scenario production cross sections times branching ratios (in fb) at Leading Order (LO) for various dark matter mass ($M_{\chi_{1}}$) and $Z^{\prime}$ mass ($M_{Z^{\prime}}$) values, using coupling constants $\texttt{g}_{l} = 0.003$ and $\texttt{g}_{DM} = 1.0$, at $\sqrt{s} = 500 \text{ GeV}$.}
\label{table:tabchi}
\end{sidewaystable}

\section{The simulation of SM backgrounds and signal samples.}
\label{section:MCandDat}
The background processes in the SM that produce muon pairs in the signal region include several interactions. These are primarily the production of Z bosons, where electron-positron pairs can lead to muon pairs ($e^{+}e^{-}$ $\rightarrow \mu^+\mu^-$). Additionally, we observe the creation of top quark pairs, specifically in the process $\text{t}\bar{\text{t}} \rightarrow \mu^+\mu^- + 2b + 2\nu$. Another significant source comes from diboson production, which encompasses several pathways: $W^{+}W^{-} \rightarrow l^+l^- + 2\nu$ ($l = \mu, e$), $ZZ \rightarrow \mu^+\mu^- + 2\nu$, and $ZZ \rightarrow 4\mu$.

The signal samples for the LV scenario, along with the corresponding SM background processes, were generated using the WHIZARD event generator, version 3.1.1 \cite{MG5}. Additionally, the effects of initial-state radiation (ISR) were taken into account and interfaced with Pythia 6.24, which is employed for the parton shower model and hadronization \cite{R34}. 

Finally, a fast detector simulation for the ILCgen detector, along with parameterizations for a generic ILC detector, was created \cite{ilcgen1, ilcgen2}. This simulation is included with the DELPHES package \cite{delphes}. 
These were generated from electron-positron collisions at the ILC with $\sqrt{s} = 500$ GeV, akin to Run 1 conditions.
\begin{table} 
\small
\centering
\begin {tabular} {|l|l|l|c|l|}
\hline
Process \hspace{1cm} & Deacy channel  & Generator  & {$\sigma \times \text{BR} ~(\text{fb})$} & Order \\
\hline
\hline
$\text{$e^{+}e^{-}$}$ & $\mu^{+}\mu^{-}$ & Whizard & 1767 & LO\hspace{6cm}\\
\hline
$\text{t}\bar{\text{t}}$ & $\mu^{+}\mu^{-} + 2b + 2\nu$ & Whizard& 10.4 & LO \\
\hline
WW & $l^{+}l^{-} + 2\nu$ & Whizard & 931 & LO \\
\hline
ZZ & $\mu^{+}\mu^{-} + 2\nu$  & Whizard & 3.7 & LO \\
\hline
ZZ  & $4\mu$ & Whizard & 0.5 & LO \\
\hline
\end {tabular}
\vspace{3pt}
\caption{Simulated SM backgrounds from electron-positron collisions at the ILC ($\sqrt{s} = 500$ GeV) are presented, including sample names, decay channel, used generators, cross-section times branching ratios, and generation order.}
\label{table:tab3}
\end{table}

For the scenario involving light vectors and dark matter particles ($\chi_{1}$ and $\chi_{2}$), we assume the mass values outlined in Table \ref{table:tab1}. With $\texttt{g}_{l} = 0.003$ and $\texttt{g}_{DM} = 1.0$, Table \ref{table:tabchi} presents the Leading Order (LO) production cross section times branching ratios for different Z$^{\prime}$ and DM mass points at the ILC with $\sqrt{s}$ = 500 GeV.

The Monte Carlo simulations have been utilized to generate the SM background samples and calculate their cross-sections in leading order, as shown in Table \ref{table:tab3}. 
The signal samples and SM background processes were estimated from these simulations, normalized to their cross-sections, and an integrated luminosity of 4 ab$^{-1}$. 

The systematic uncertainties considered in this analysis include the following: integrated luminosity contributes 0.26\%, while the shape of the luminosity spectra adds a significant 5\%. Additionally, the beam polarization is estimated to be between 0.02\% and 0.08\%, and the beam energy uncertainty is pegged at 0.2\%. Furthermore, the Delphes detector performance, which encompasses tracking and muon energy resolution, introduces about 4\% uncertainty. When we take all these factors into account, an overall flat 10\% uncertainty was applied to account for systematic effects \cite{monoPhoton-ilc1,monoPhoton-ilc2,delphes-systematics}. 

\section{Event selection}
\label{section:AnSelection}
The event selection process has been carefully crafted to identify a final state characterized by two muons with low transverse momentum $(p_{T})$ and missing transverse energy, which indicates the presence of a dark matter candidate. 
To achieve this, a series of criteria are applied to various kinematic parameters. 

Both muons are required to undergo a preliminary selection, which includes the following criteria: \\
- $p^{\mu}_{T}$ (GeV) $> 10$, since we are studying the angular distributions for Z$^{\prime}$ with a mass range from 20 GeV to 100 GeV. Therefore, we have chosen the transverse momentum ($p^{\mu}_{T}$) of the muon to be greater than 10 GeV.\\
- $|\eta^{\mu}|$ (rad) $<$ 2.5,\\
- $\text{IsolationVar} < 0.1$. 

In DELPHES software, the term $\text{"IsolationVar"}$ refers to the isolation cut used to filter out muons that originate inside jets. This criterion stipulates that the sum of the scalar $p_{T}$ of all muon tracks within a cone of $\Delta R = 0.5$ around the muon candidate, excluding the candidate itself, must not exceed 10\% of the $p_{T}$ of the muon. 

The same kinematic criteria are also utilized for the pre-selection of electrons. In addition to these, we introduced another criterion that examines the ratio of energy deposited in the hadronic calorimeter ($E_{had}$) to that in the electromagnetic calorimeter ($E_{em}$). This ratio should be less than 5\%.

Each event is selected based on the presence of two oppositely charged muons, which are used to evaluate the dimuon invariant mass. Additionally, we also consider events with one positively charged electron and one negatively charged muon for the $e^+\mu^-$ mass spectrum, which will be discussed in section \ref{section:BkgReduction}. In both cases, the invariant mass must exceed 10 GeV.

\begin{figure}
\centering
\subfigure[]{
  \includegraphics[width=74.0mm]{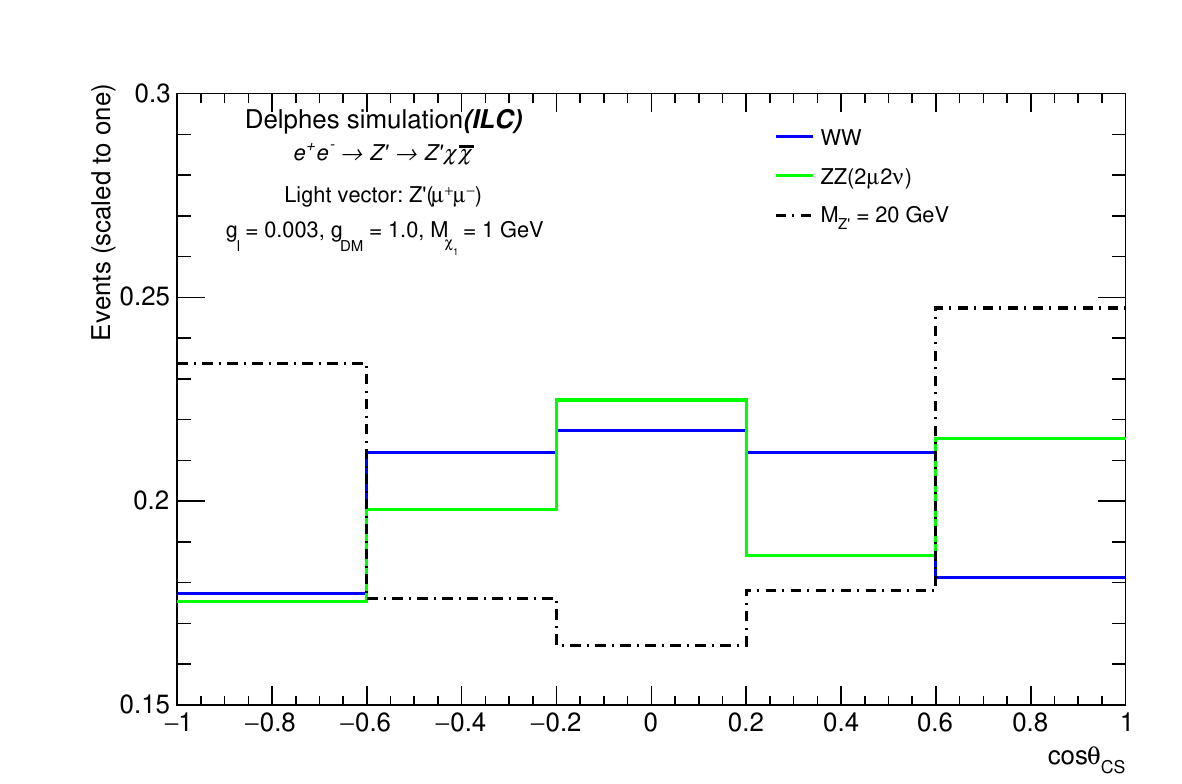}
  \label{cos1}
}
\subfigure[]{
  \includegraphics[width=74.0mm]{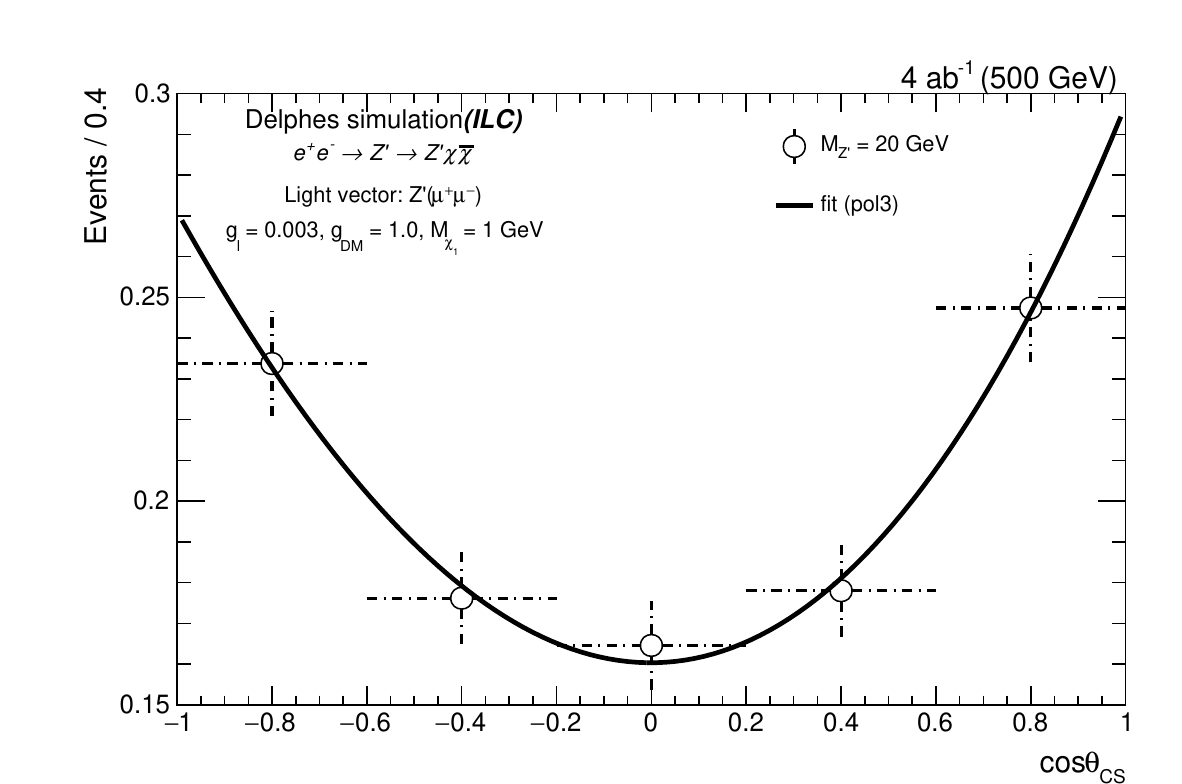}
  \label{cos2}
}
\caption{Normalized cos$\theta_{CS}$ distributions for a resonant model in the LV scenario with a $Z^{\prime}$ mass of 20 GeV, analyzing WW and ZZ($2\mu 2\nu$) events at $\sqrt{s} = 500$ GeV in \ref{cos1},
and for the LV signal, only a polynomial 3 function fit is applied as shown in \ref{cos2}. 
Events must meet pre-selection criteria from Table \ref{cuts} and have a reconstructed invariant mass between 18 - 45 GeV. Histograms are normalized to unity to emphasize qualitative features.}
\label{figCS}
\end{figure}

\begin{table}
\small
    \centering
    \begin{tabular}{|c|c|c|}
\hline
Pre-selection ($\mu$) & Pre-selection ($e$)  \\
\hline
    \hline
  $p^{\mu}_{T} >$ 10 GeV &  $p^{e}_{T} >$ 10 GeV \\
  $|\eta^{\mu}| <$ 2.5 rad   & $|\eta^{e}| <$ 2.5 rad \\
  $\Sigma_{i} p^{i}_{T}/p^{\mu}_{T} < 0.1$ &  $\Sigma_{i} p^{i}_{T}/p^{e}_{T} < 0.1$\\
    & $E_{had}/E_{em} <$ 0.05  \\
    $M_{\mu^{+}\mu^{-}} > 10$ GeV&$M_{e^{+}\mu^{-}} > 10$ GeV\\
    \hline
    \end{tabular}
    \caption{Summary of cut-based event pre-selections used in the analysis for muons and electrons.}
    \label{pre-cuts}
\end{table}

As seen from Figure \ref{figure:fig1}, there are two Z$^{\prime}$ bosons present. The first Z$^{\prime}$ acts as a mediator (off-shell) particle, which decays into a dark matter (DM) particle and an anti-dark matter particle. The DM can then radiate the second, on-shell Z$^{\prime}$ boson, which subsequently decays into a dimuon pair. 
The angular distribution of the decay products ($\mu^{+} \mu^{-}$) can be expressed in terms of the Collins-Soper angle. This distribution is sensitive to the spin of the Z$^{\prime}$ boson.

We present the cos$\theta_{CS}$ distribution for a resonant model grounded in the light vector scenario, as illustrated in Figure \ref{figCS}. The mass of dark boson (Z$^{\prime}$) was chosen with a low mass of 20 GeV, considered as a benchmark point due to its optimal cross-section times branching ratio, as indicated in Table \ref{table:tabchi}. We also compare these results with the irreducible backgrounds from WW and ZZ($2\mu 2\nu$) events in Figure \ref{cos1}.
The plot presented in Figure \ref{cos2} shows only the LV signal fitted by a polynomial function of order 3.
All events adhere to the pre-selection criteria detailed in Table \ref{pre-cuts} and exhibit a reconstructed invariant mass ranging from 18 to 45 GeV.
The results are illustrated with a dotted line representing the model signal and blue and green lines for the WW and ZZ events, normalized to unity. We observe a clear distinction between the simplified model and the WW and ZZ events. 

The signal shape, presented in Figure \ref{cos2}, exhibits a typical characteristic of a spin-1 boson, displaying a symmetric distribution around zero. This distribution aligns with the findings from the study conducted in \cite{Osland}.
In addition, research conducted by the ATLAS collaboration has demonstrated that the cos$\theta_{CS}$ distribution can effectively differentiate between a scalar boson (with spin-0) and a newly proposed vector boson, the Randall–Sundrum models (RS) graviton (with spin-2) \cite{graviton}.
\begin{figure}
\small
\centering
\resizebox*{9.2cm}{!}{\includegraphics{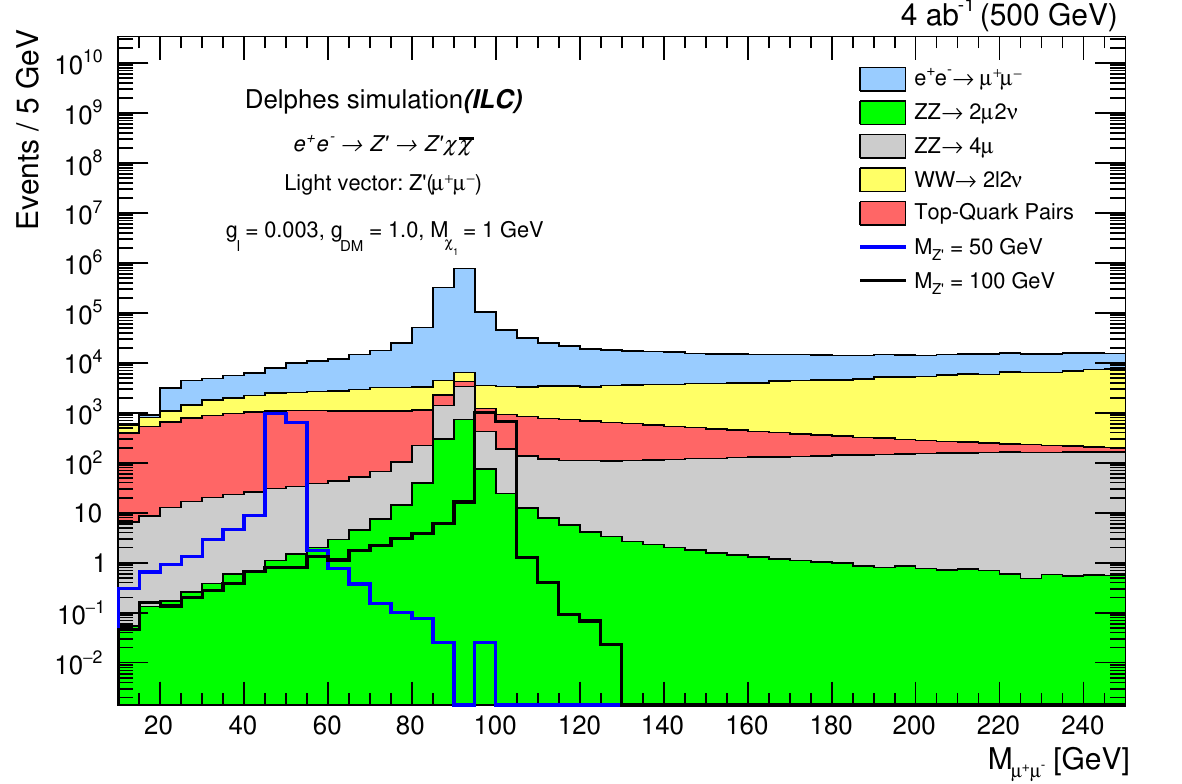}}
\caption{The dimuon invariant mass spectrum, after pre-selection (see Table \ref{pre-cuts}), for estimated SM backgrounds and various neutral gauge boson (Z$^{\prime}$) masses based on the LV scenario, with dark matter mass ($M_{\chi_{1}} = 1$ GeV).}
\label{figure:fig3}
\end{figure}

In Figure \ref{figure:fig3}, the invariant mass spectrum for dimuon events is presented; these events meet the pre-selection criteria outlined in Table \ref{pre-cuts}. 
In this figure, the cyan histogram illustrates the Z background resulting from the interaction $e^{+}e^{-} \rightarrow \mu^{+}\mu^{-}$. On the other hand, the various diboson backgrounds are depicted in distinct colors: yellow represents WW, gray indicates $ZZ(4\mu)$, and green corresponds to $ZZ(2\mu2\nu)$.
The red histogram indicates the $t\bar{t}$ background. These histograms are displayed in a stacked format, allowing for easy comparison. The signals for the LV scenario are shown for medium (50 GeV) and large (100 GeV) mass points of the Z$^{\prime}$ boson, with the dark matter mass fixed at $M_{\chi_{1}} = 1$ GeV. Different colored lines represent these signals, which are overlaid on the same graph.
\begin{figure*}
\centering
\subfigure[]{
  \includegraphics[width=73mm]{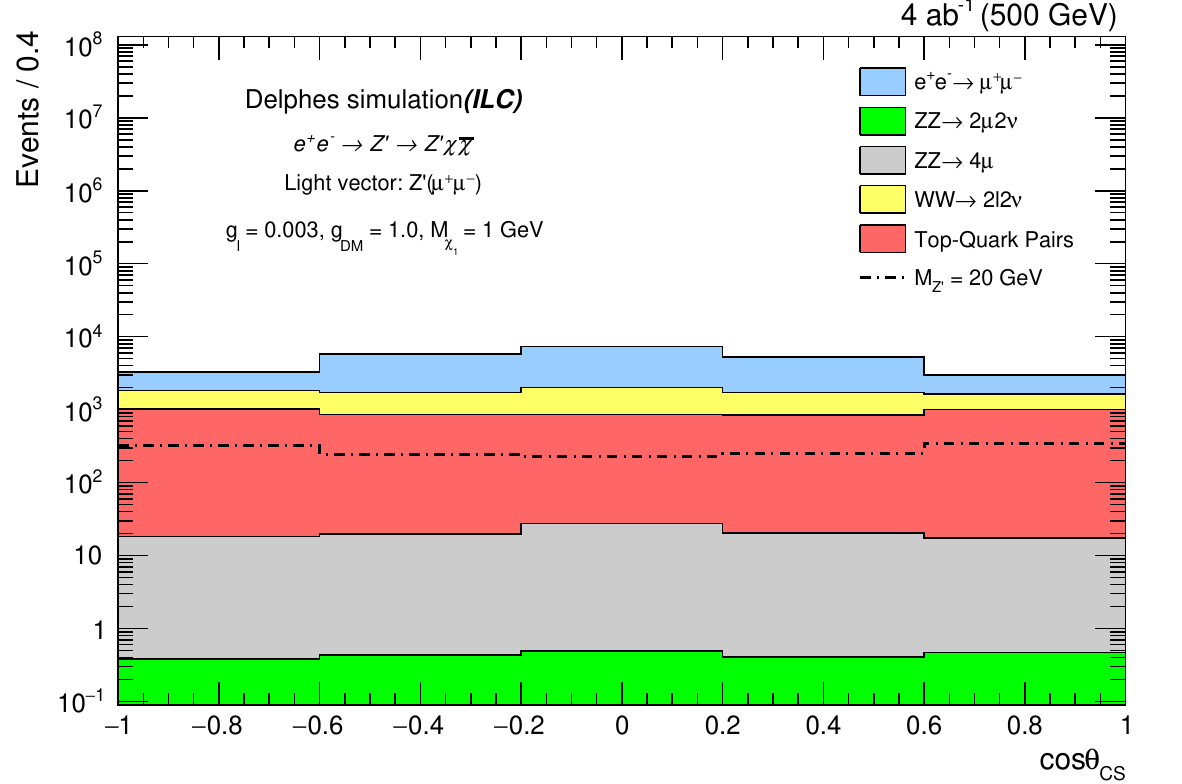}
  \label{bin20}
}
\subfigure[]{
  \includegraphics[width=73mm]{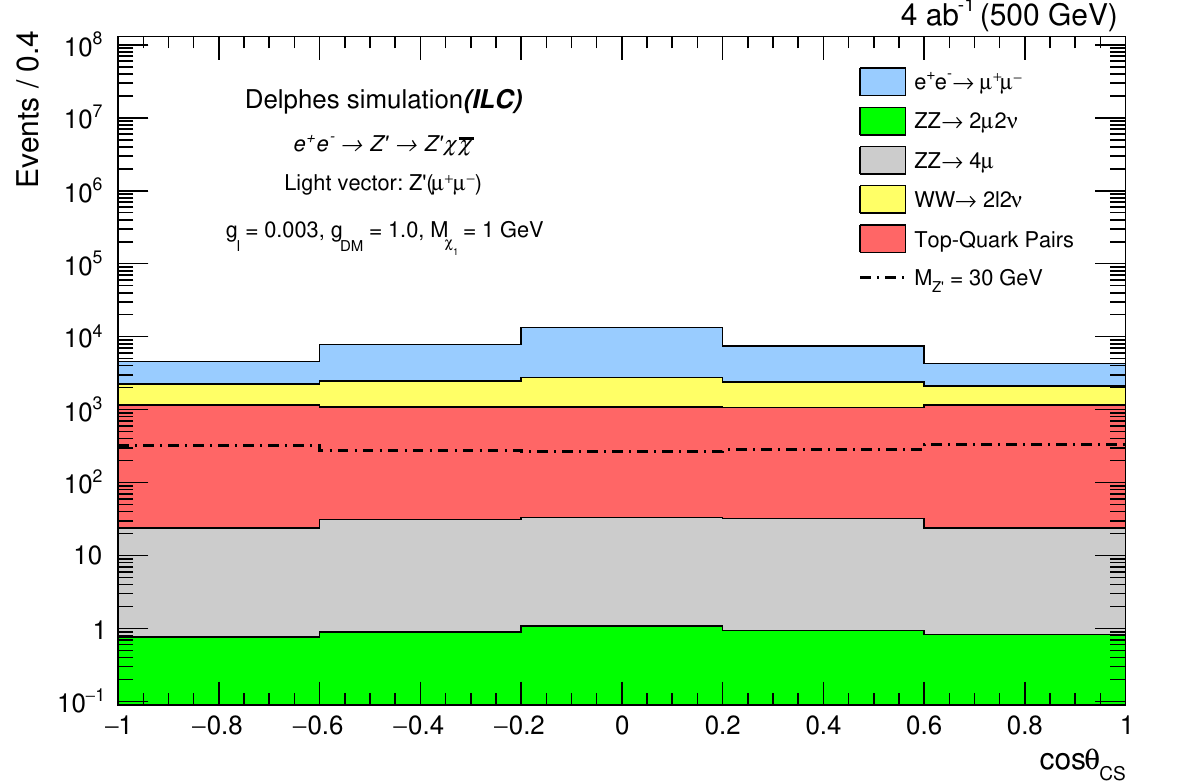}
  \label{bin30}
}
\subfigure[]{
  \includegraphics[width=73mm]{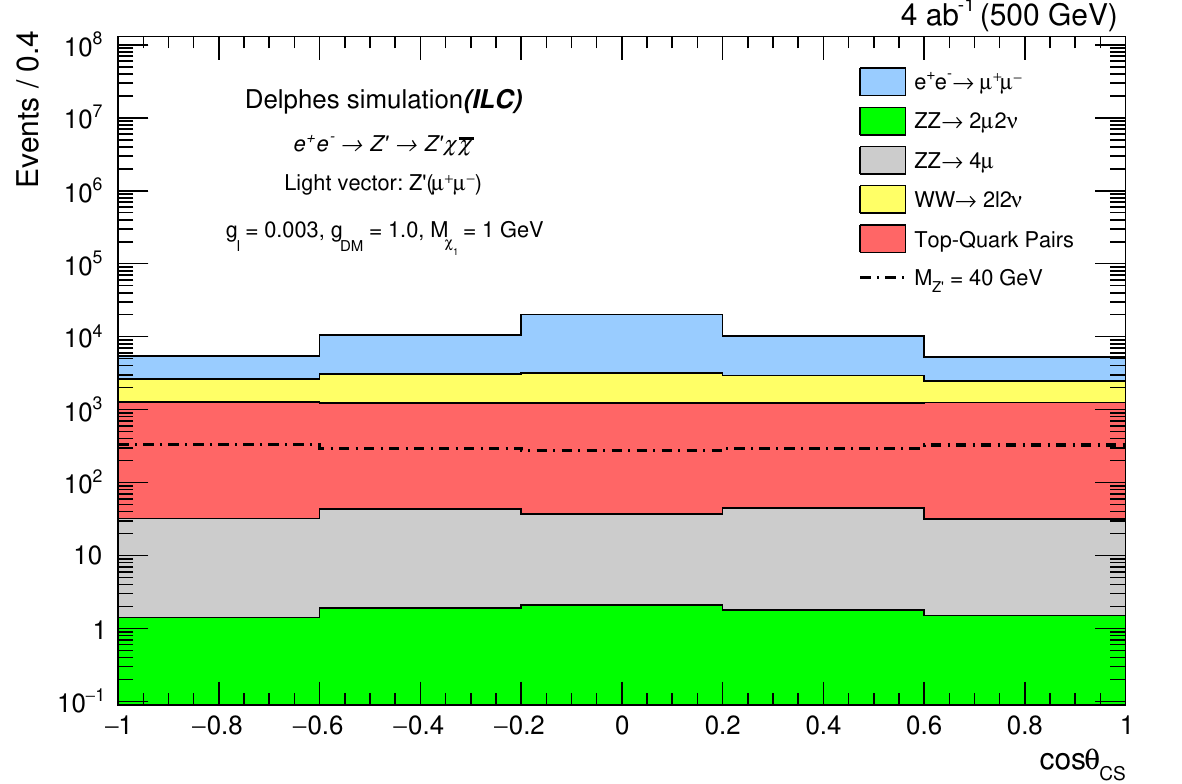}
  \label{bin40}
}
\subfigure[]{
  \includegraphics[width=73mm]{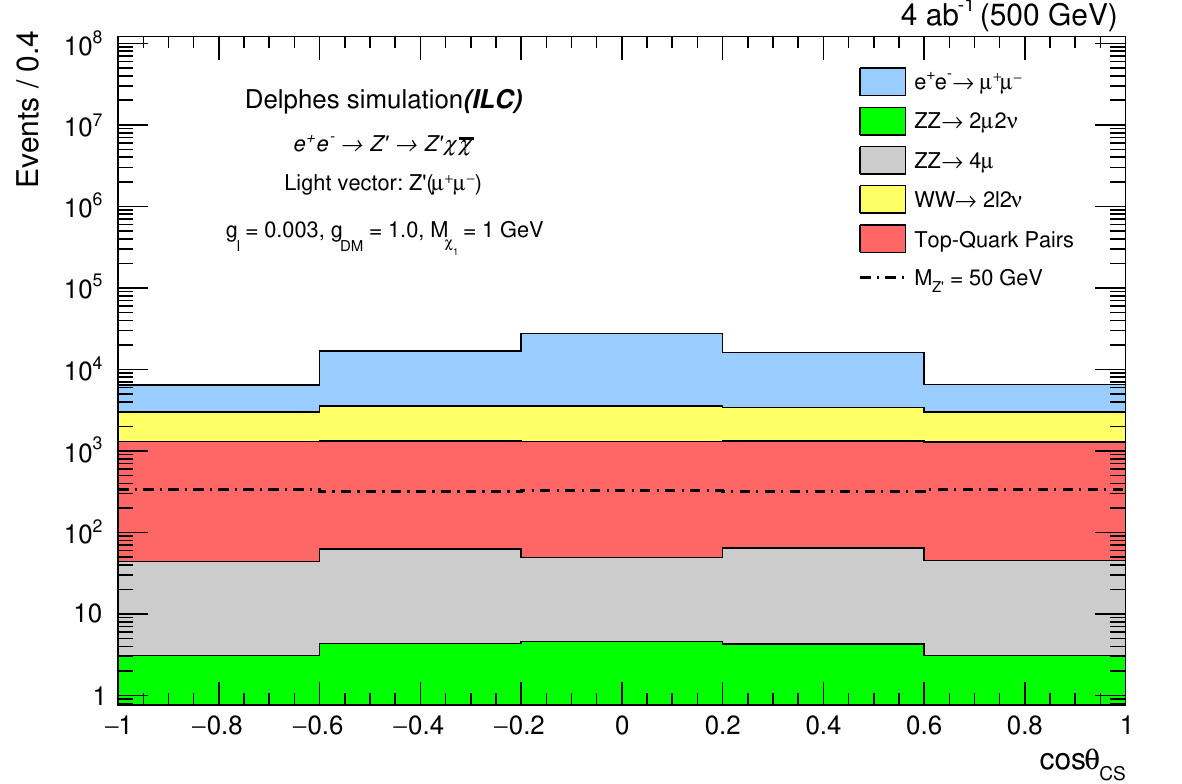}
  \label{bin50}
}
\subfigure[]{
  \includegraphics[width=73mm]{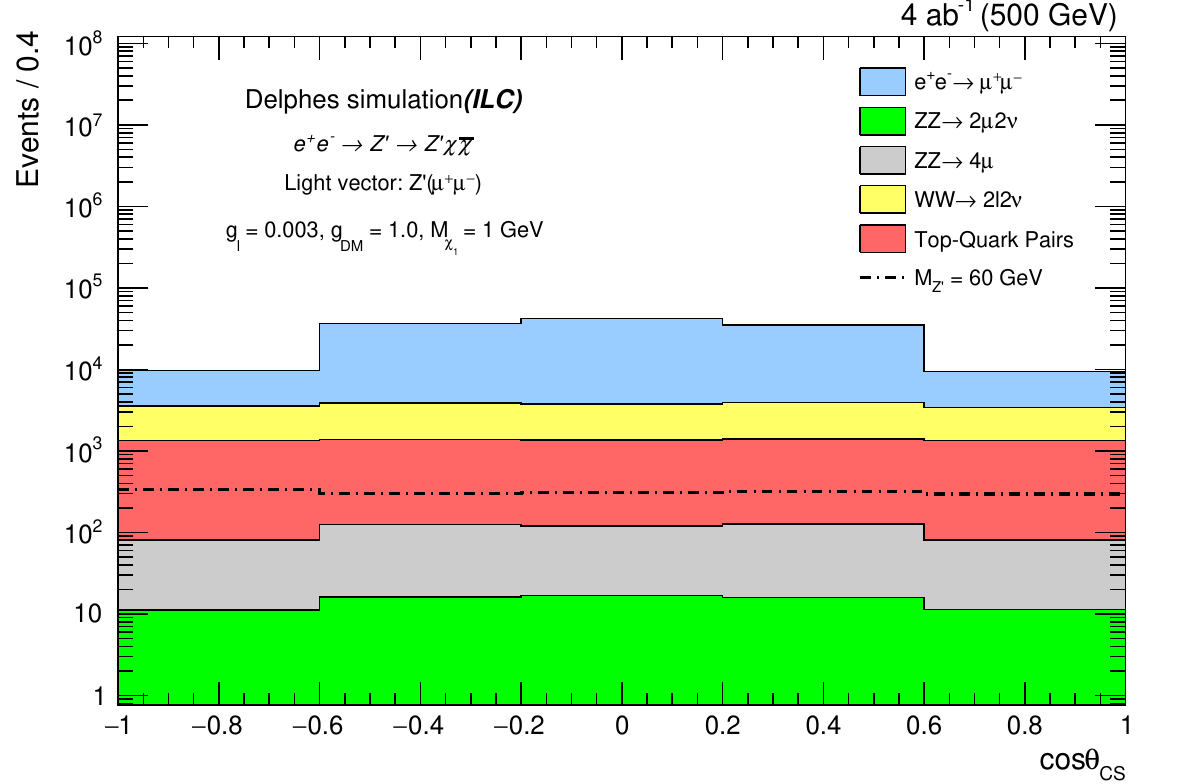}
  \label{bin60}
}

\caption{The distributions of cos$\theta_{CS}$ are presented for events that pass the pre-selection criteria listed in Table \ref{pre-cuts}. The histograms show the standard model expectations, while the signal samples corresponding to the light vector model with different mass values $M_{A^{\prime}}$ ranging from 20 to 60 GeV are also superimposed.
The analysis focuses on several dimuon mass windows, specifically:
18 $< M_{\mu^+\mu^-} <$ 45 GeV \ref{bin20},
27 $< M_{\mu^+\mu^-} <$ 55 GeV \ref{bin30},
36 $< M_{\mu^+\mu^-} <$ 65 GeV \ref{bin40},
45 $< M_{\mu^+\mu^-} <$ 75 GeV \ref{bin50},
54 $< M_{\mu^+\mu^-} <$ 85 GeV \ref{bin60}. 
}
\label{costhetaBeforecuts}
\end{figure*}

The distributions of \(\cos\theta_{CS}\) are presented in Figure \ref{costhetaBeforecuts} for events that meet the pre-selection criteria outlined in Table \ref{pre-cuts}. The histograms depict the expected outcomes from the standard model, alongside signal samples from the light vector model for different mass values \(M_{A^{\prime}}\) ranging from 20 to 60 GeV. This analysis zeroes in on various dimuon mass windows, specifically:
18 $< M_{\mu^+\mu^-} <$ 45 GeV \ref{bin20}, 27 $< M_{\mu^+\mu^-} <$ 55 GeV \ref{bin30},
36 $< M_{\mu^+\mu^-} <$ 65 GeV \ref{bin40}, 45 $< M_{\mu^+\mu^-} <$ 75 GeV \ref{bin50},
54 $< M_{\mu^+\mu^-} <$ 85 GeV \ref{bin60}. 

\section{Background reduction}
\label{section:BkgReduction}
The invariant mass plots shown in Figure \ref{figure:fig3} and the $\cos\theta_{CS}$ distribution in Figure \ref{costhetaBeforecuts} reveal that the signal samples are significantly mixed with background events throughout the entire dimuon invariant mass range. Therefore, it is crucial to adopt more stringent methods and criteria to differentiate the signals from the standard model backgrounds effectively.
\subsection{$e\mu$ method}
\label{section:emu}
The invariant mass plots shown in Figure \ref{figure:fig3} reveal that the second-largest contribution to the SM background originates from the WW background. This particular background can be effectively diminished by subtracting the contributions from $e^+\mu^-$.
\begin{figure}
\centering
\subfigure[]{
  \includegraphics[width=73.0mm]{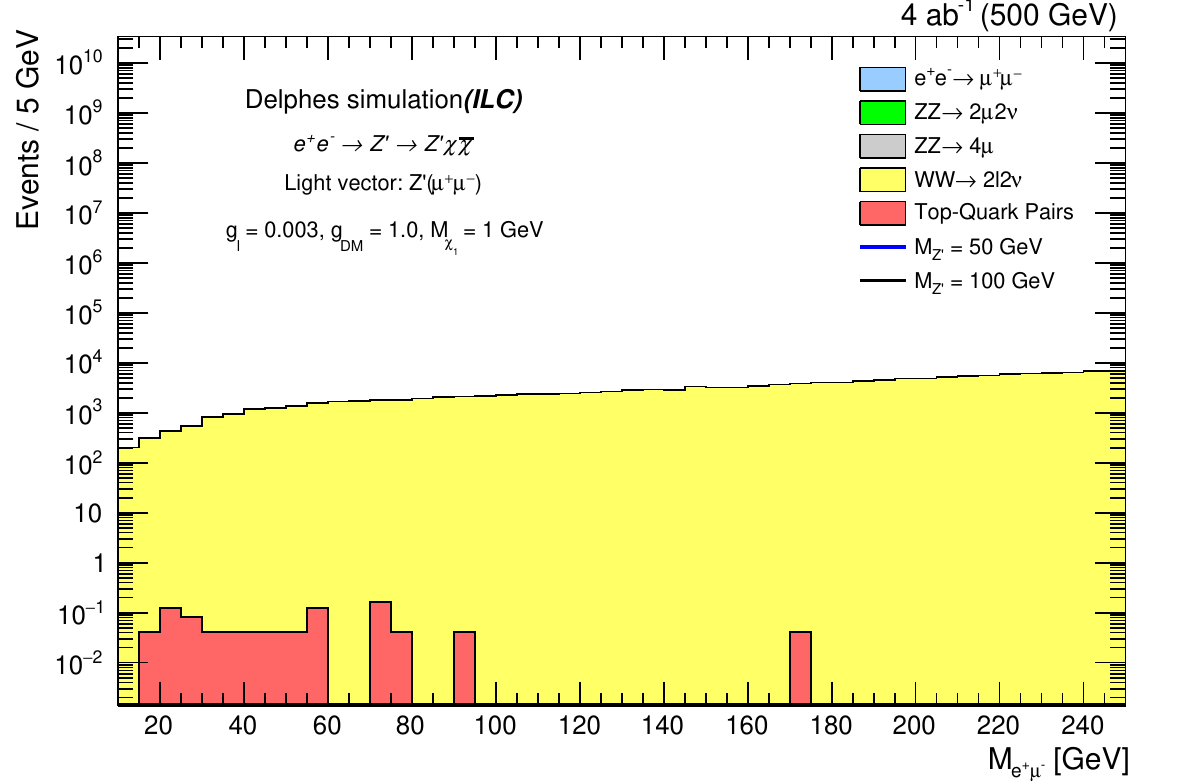}
  \label{mass-emu}
}
\subfigure[]{
  \includegraphics[width=73.0mm]{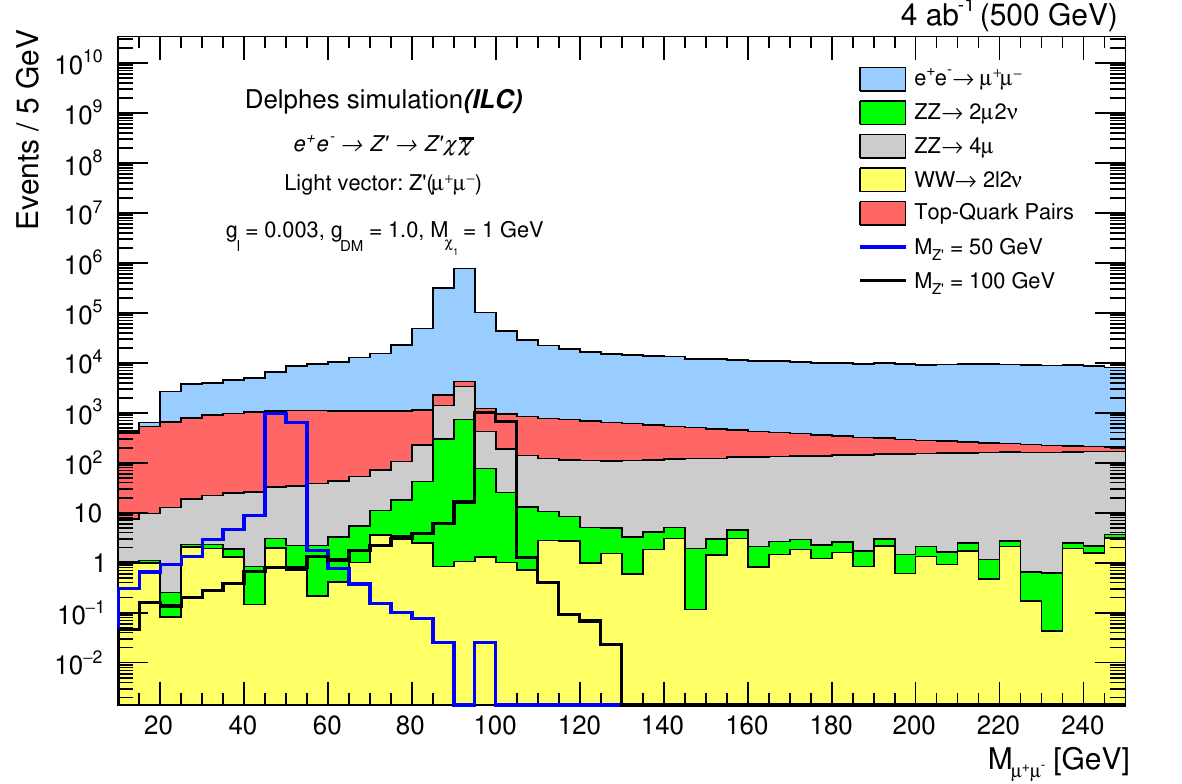}
  \label{mass-residual}
}
\caption{The invariant mass spectrum for $e^+\mu^-$ events in \ref{mass-emu}, and for dimuon events after subtraction of $e^+\mu^-$ events in \ref{mass-residual} that pass the pre-selection listed in Table \ref{pre-cuts}. 
For the estimated SM backgrounds and various neutral gauge boson (Z$^{\prime}$) masses based on the LV scenario, with dark matter mass ($M_{\chi_{1}} = 1$ GeV).}
\label{figure:mass}
\end{figure}

We present the mass spectrum for \( e^+\mu^- \) events, as shown in Figure \ref{mass-emu}. Additionally, we show the spectrum for dimuon events after subtracting the contributions from \( e^+\mu^- \) events, as illustrated in Figure \ref{mass-residual}. 
As a result of this subtraction, Figure \ref{mass-residual} displays the spectrum for dimuon events, which significantly reduces the contamination from the \( WW \) background.


\subsection{Event final selection and efficiencies}
\label{section:finalcute}
\begin{figure*}
\centering
\subfigure[]{
  \includegraphics[width=73.0mm]{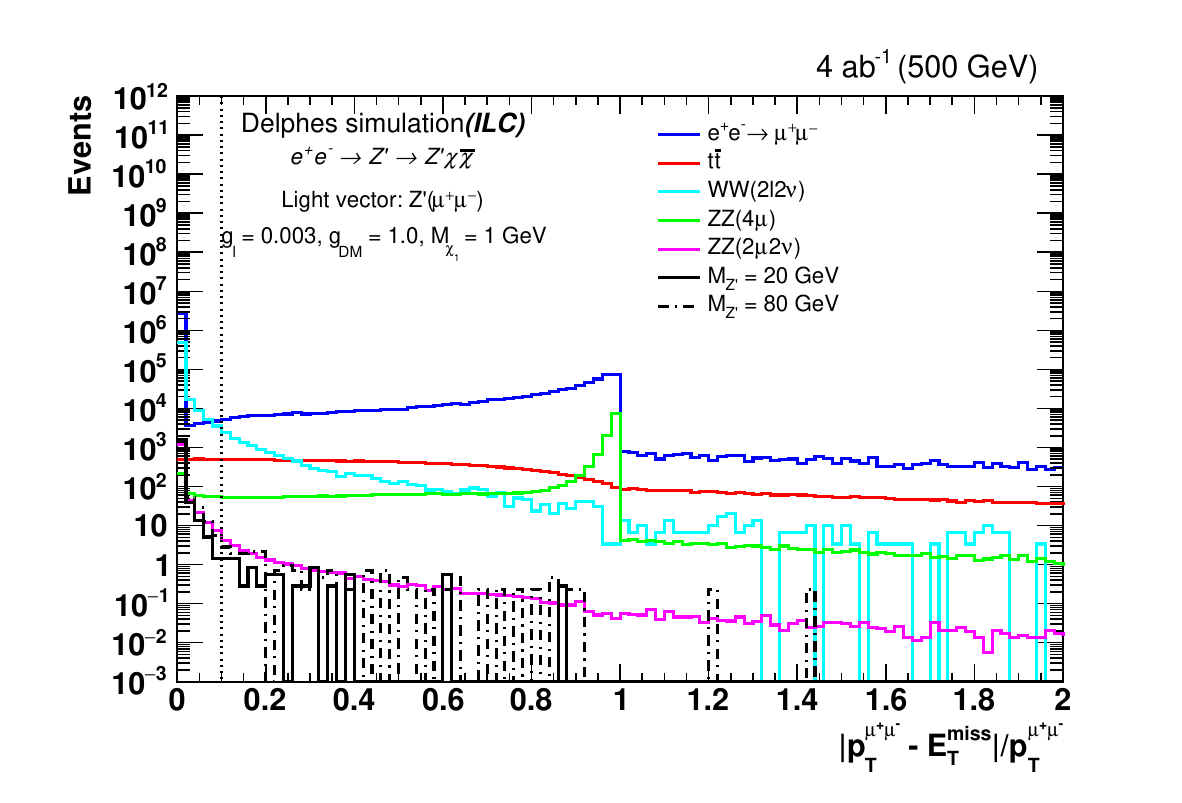}
  \label{figure:ptdiff}
}
\subfigure[]{
  \includegraphics[width=73.0mm]{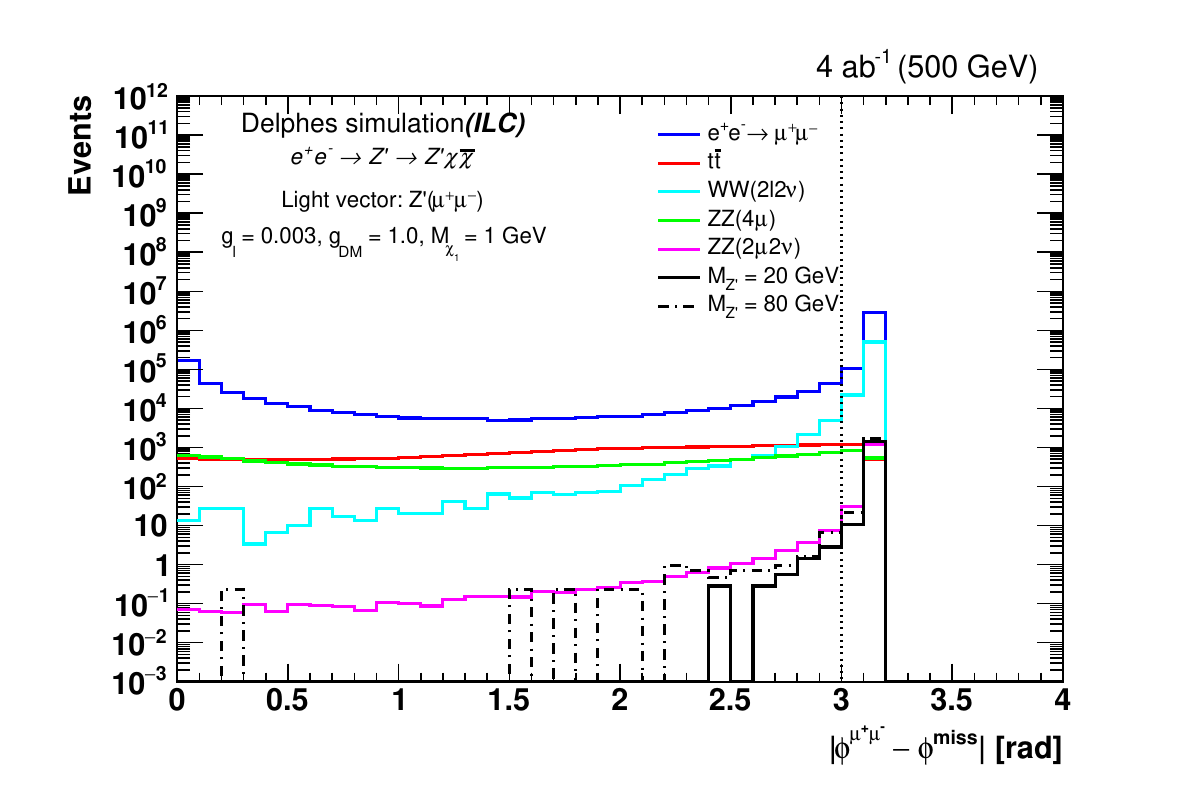}
  \label{figure:deltaphi}
}
\subfigure[]{
  \includegraphics[width=73.0mm]{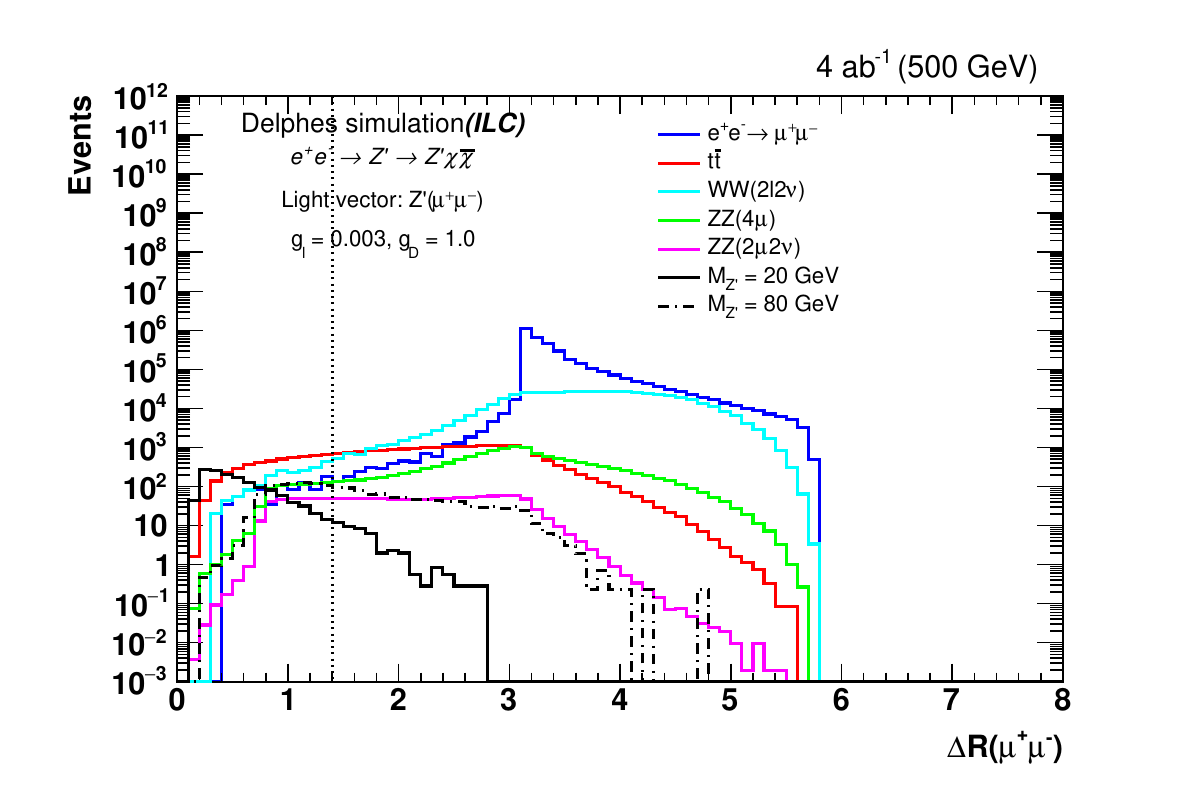}
  \label{figure:deltar}
}
\subfigure[]{
  \includegraphics[width=73.0mm]{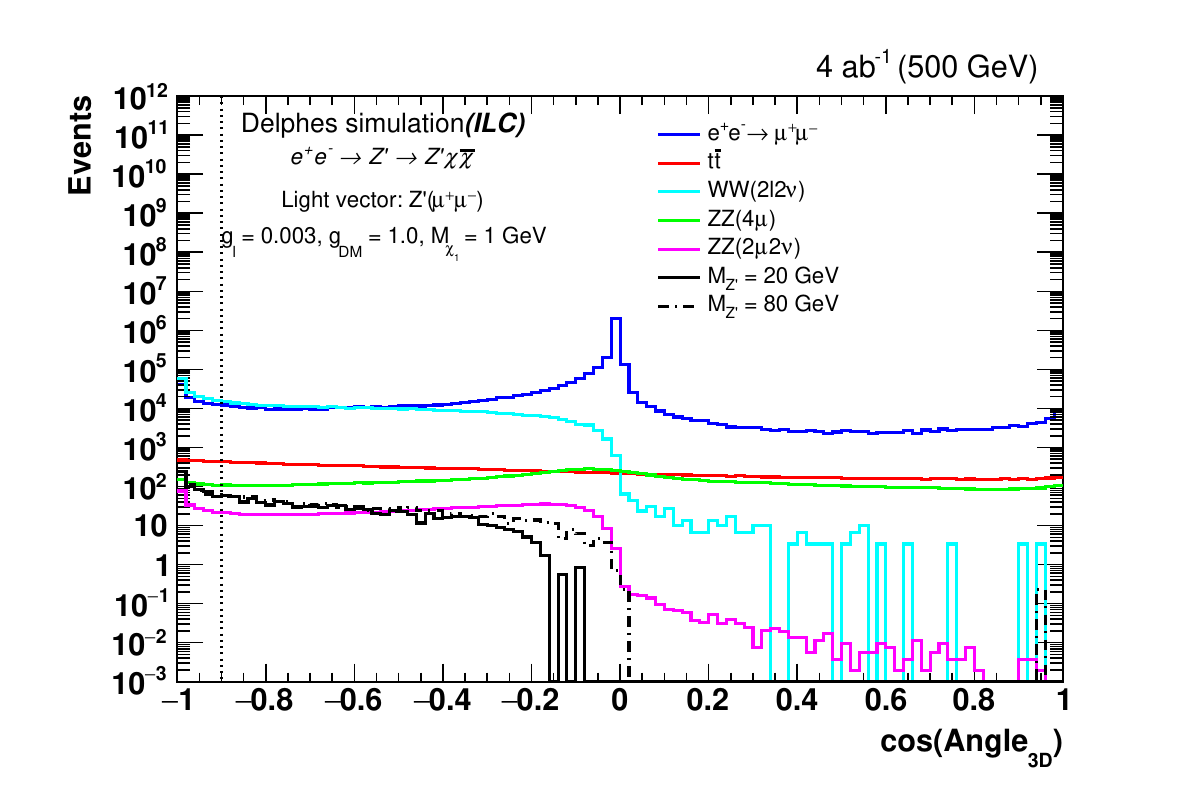}
  \label{figure:3Dangle}
}
\subfigure[]{
  \includegraphics[width=73.0mm]{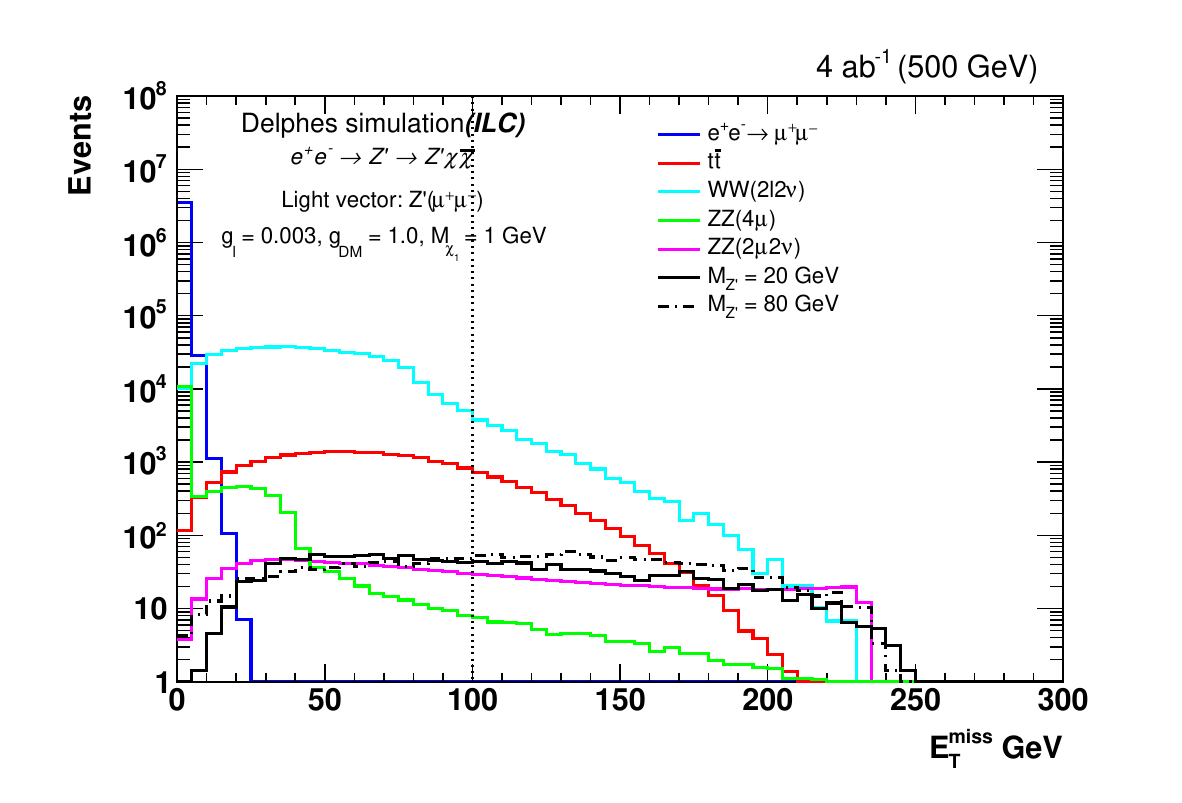}
  \label{figure:etmiss}
}
\caption{The distributions of extra variables for dimuon events are presented, where both muons meet the low \( p_T \) ID criteria from Table \ref{pre-cuts}. The five examined variables are: \( |p_{T}^{\mu^{+}\mu^{-}} - E_{T}^{miss}| / p_{T}^{\mu^{+}\mu^{-}} \) (see \ref{figure:ptdiff}),
\( \Delta\phi_{\mu^{+}\mu^{-},\vec{E}_{T}^{miss}} \) (see \ref{figure:deltaphi}),
\( \Delta R(\mu^{+}\mu^{-}) \) (see \ref{figure:deltar}),
\( \text{cos}(\text{Angle}_{3D}) \) (see \ref{figure:3Dangle}), and
\( E_{T}^{\text{miss}} \) (see \ref{figure:etmiss}).
The model explores the LV scenario with two \( Z^{\prime} \) masses (20 and 80 GeV) alongside SM backgrounds. Vertical dashed lines indicate the cut values for each variable.}
\label{figure:fig70}
\end{figure*}
In addition to the pre-selection criteria, we have applied tighter cuts based on six variables:

1. We restrict the invariant mass of the dimuon to a narrow range around the mass of the neutral gauge boson \( Z^{\prime} \). Specifically, we require that \( 0.9 \times M_{Z^{\prime}} < M_{\mu^{+}\mu^{-}} < M_{Z^{\prime}} + 25 \), as suggested in reference \cite{R1}.

2. We assess the relative difference between the transverse momentum of the dimuon \( (p_{T}^{\mu^{+}\mu^{-}}) \) and the missing transverse energy \( (E^{\text{miss}}_{T}) \). This difference is selected to be less than 0.1, defined by the condition \( |p_{T}^{\mu^{+}\mu^{-}} - E^{\text{miss}}_{T}|/p_{T}^{\mu^{+}\mu^{-}} < 0.1 \).

3. We calculate the azimuthal angle difference \( \Delta\phi_{\mu^{+}\mu^{-},\vec{E}^{\text{miss}}_{T}} \), which is the difference between the azimuthal angles of the dimuon and the missing transverse energy \( (|\phi^{\mu^{+}\mu^{-}} - \phi^{\text{miss}}|) \). This value is required to be greater than 3.0 radians.

4. We examine the angular separation in $\eta$ and $\phi$ coordinates \( \Delta R(\mu^{+}\mu^{-}) \) between the two opposite-sign muons, which must be less than 1.4.

5. We apply a criterion on the cosine of the 3D angle between the missing energy vector and the dimuon system vector to ensure they are back-to-back, requiring that \( \cos(\text{Angle}_{3D}) < -0.9 \).

6. Finally, we impose a cut on the missing transverse energy, requiring that \(E_{T}^{miss} > 100\) GeV.

Table \ref{cuts} outlines the final analysis event selection presented in the paper. This stage incorporates the muon pre-selection cuts, along with an additional six tight cuts, as explained above.

The graphs presented in Figure \ref{figure:fig70} display the distributions of certain variables for two signal presentations ($M_{Z^{\prime}} =$ 20 and 80 GeV) from the simplified model related to the LV scenario. These variables are compared with the SM backgrounds for dimuon events that meet the pre-selection criteria outlined in Table \ref{pre-cuts}.

The first variable is represented as \( |p_{T}^{\mu^{+}\mu^{-}} - E_{T}^{\text{miss}}| / p_{T}^{\mu^{+}\mu^{-}} \), and its graph is displayed in Figure \ref{figure:ptdiff}. The second variable is denoted as \( \Delta\phi_{\mu^{+}\mu^{-},\vec{E}_{T}^{\text{miss}}} \), with its corresponding graph shown in Figure \ref{figure:deltaphi}. The third variable measures the angular distance between the two muons and is referred to as \( \Delta R(\mu^{+}\mu^{-}) \), which is presented in Figure \ref{figure:deltar}. The fourth variable, \( \text{cos}(\text{Angle}_{3D}) \), is illustrated in Figure \ref{figure:3Dangle}. Finally, the fifth variable, the missing transverse energy \( E_{T}^{\text{miss}} \), is presented in Figure \ref{figure:etmiss}.
The Figures presented here depict the SM backgrounds and signals from the LV scenario. These signals are generated with a neutral gauge boson mass of \( M_{Z^{\prime}} = 20 \, \text{GeV} \) and \( 80 \, \text{GeV} \), while the dark matter mass is set at \( M_{\chi_{1}} = 1 \, \text{GeV} \). The vertical black dashed lines in these Figures indicate the selected cut values for each variable.

\begin{figure*}
\centering
\subfigure[]{
  \includegraphics[width=73.0mm]{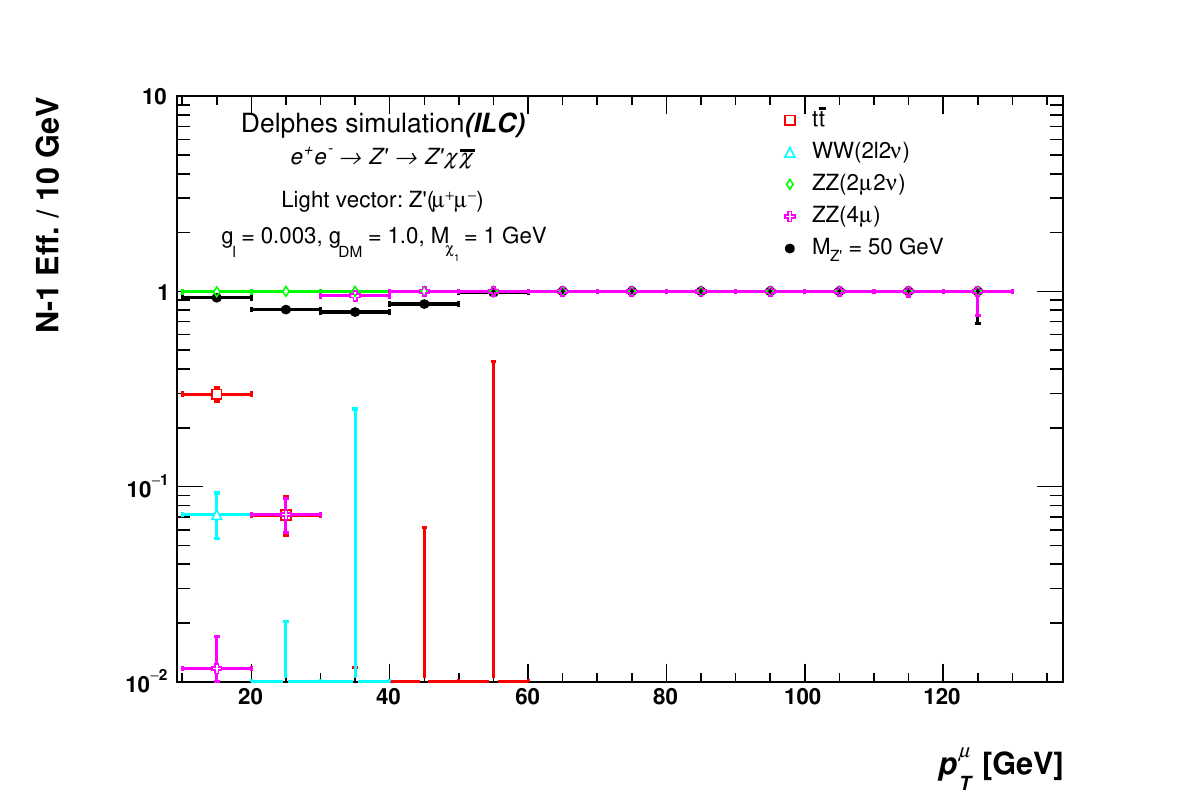}
  \label{eff1}
}
\subfigure[]{
  \includegraphics[width=73.0mm]{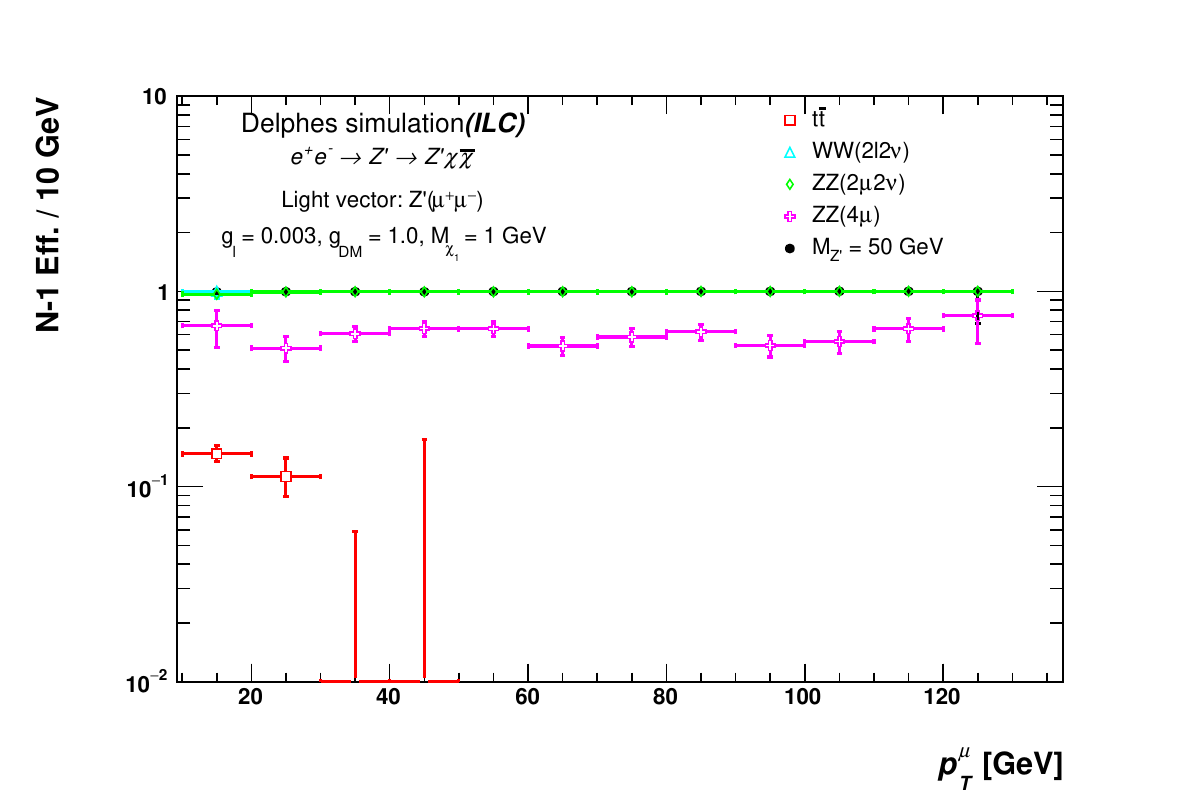}
  \label{eff2}
}
\subfigure[]{
  \includegraphics[width=73.0mm]{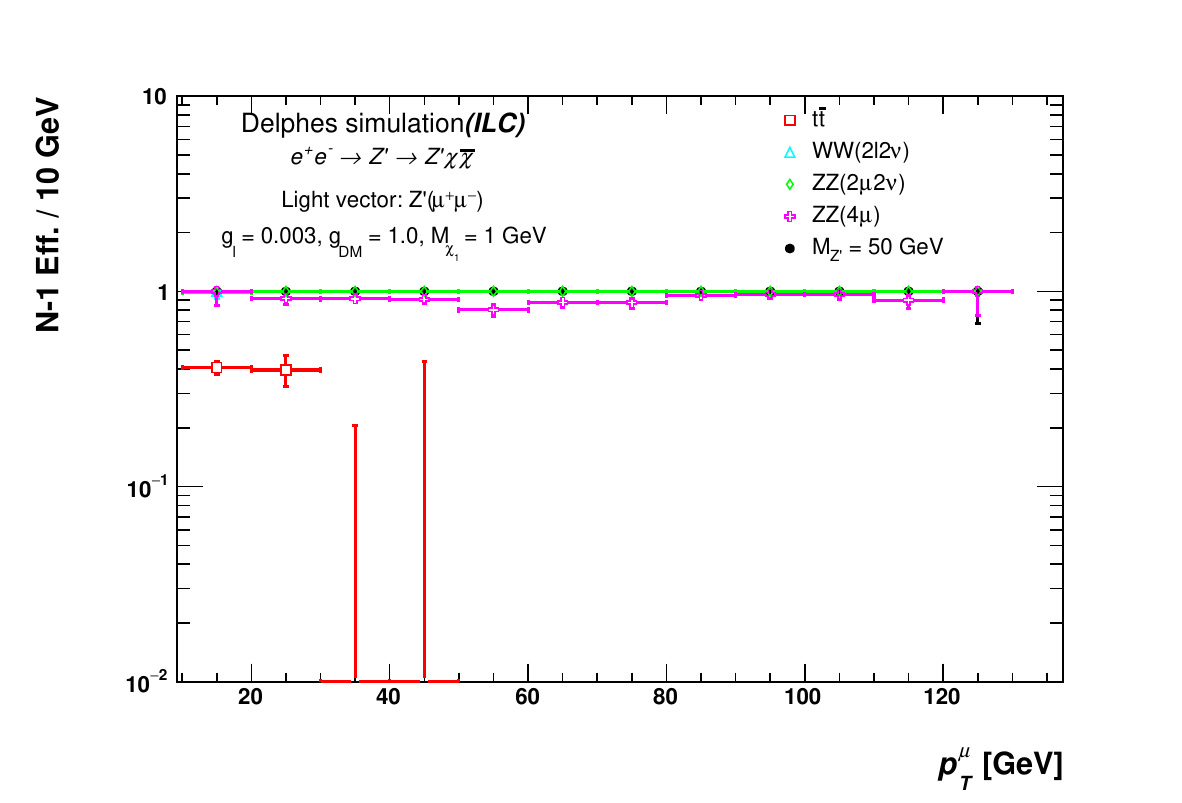}
  \label{eff3}
}
\subfigure[]{
  \includegraphics[width=73.0mm]{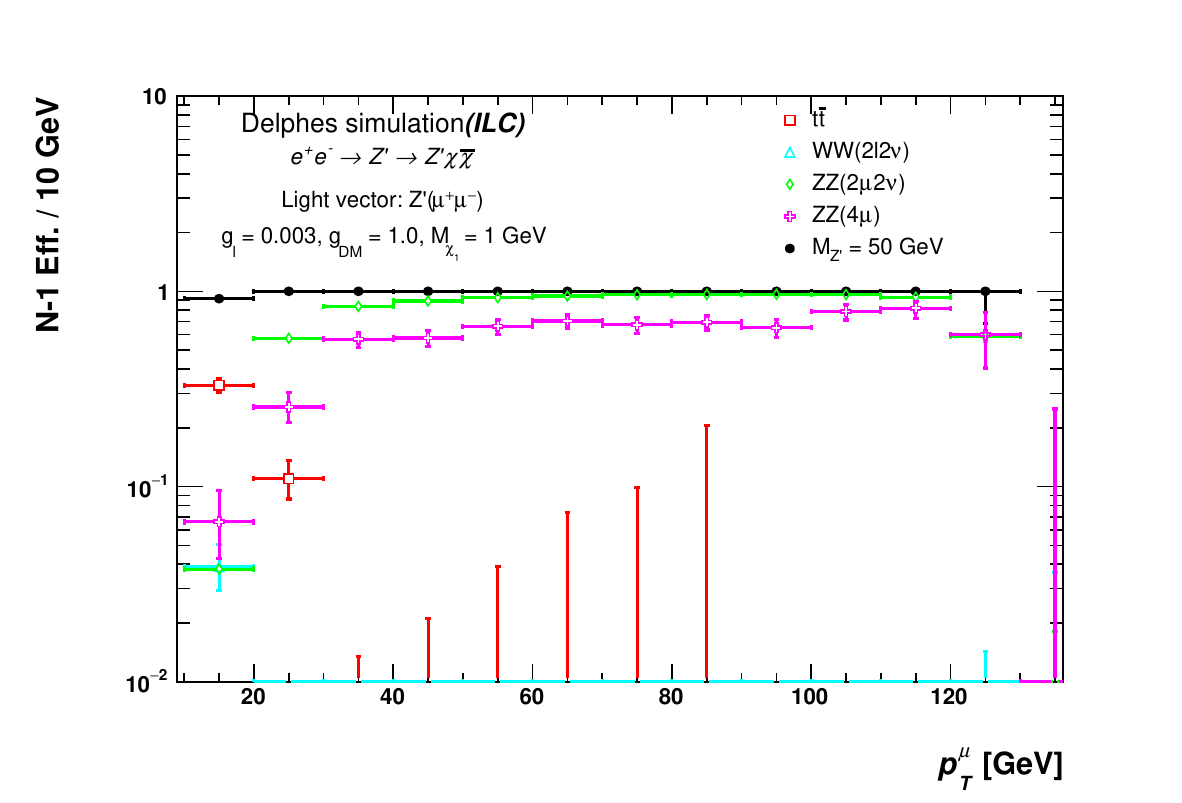}
  \label{eff4}
}
\subfigure[]{
  \includegraphics[width=73.0mm]{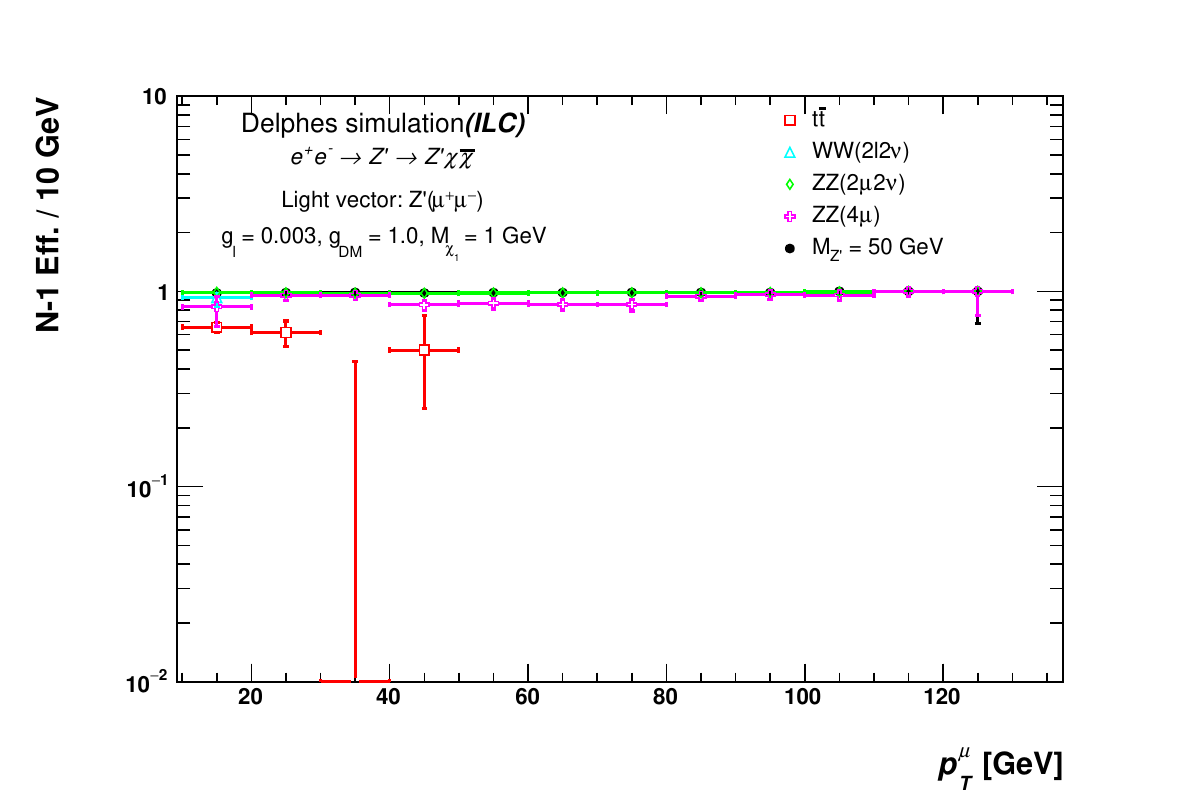}
  \label{eff5}
}

\caption{Distributions of the N-1 efficiencies plotted against the transverse momentum of the leading reconstructed muon ($p^{\mu}_{T}$) for the following cuts; 
for $E_{T}^{miss} > 100$ GeV \ref{eff1}, $|p_{T}^{\mu^{+}\mu^{-}} - E_{T}^{miss}|/p_{T}^{\mu^{+}\mu^{-}} < 0.1$ \ref{eff2}, 
$\Delta\phi_{\mu^{+}\mu^{-},\vec{E_{T}}^{\text{miss}}} > 3.0$ \ref{eff3},
\( \Delta R(\mu^{+}\mu^{-}) < 1.4\) \ref{eff4}, and
\( \text{cos}(\text{Angle}_{3D}) < -0.9\) \ref{eff5}
for the signal in the LV scenario with $M_{Z^{\prime}}= 50$ GeV and for the SM backgrounds.}
\label{Effs}
\end{figure*}

The performance metrics for fine-tuning these rigorous cuts are illustrated by plotting the N-1 efficiency for each of the five criteria detailed in Table \ref{cuts}. To calculate the N-1 efficiency, we take the number of events that successfully pass the final selection and divide it by the number of events that would have cleared the final selection in the absence of the specific cut under consideration.

In Figure \ref{Effs}, we present the distributions of the N-1 efficiencies plotted against the transverse momentum of the leading reconstructed muon (\( p^{\mu}_{T} \)) for the following conditions: 
$E_{T}^{miss} > 100$ GeV Figure \ref{eff1}, $|p_{T}^{\mu^{+}\mu^{-}} - E_{T}^{miss}|/p_{T}^{\mu^{+}\mu^{-}} < 0.1$ Figure \ref{eff2}, 
$\Delta\phi_{\mu^{+}\mu^{-},\vec{E_{T}}^{\text{miss}}} > 3.0$ Figure \ref{eff3},
\( \Delta R(\mu^{+}\mu^{-}) < 1.4\) Figure \ref{eff4}, and
\( \text{cos}(\text{Angle}_{3D}) < -0.9\) Figure \ref{eff5}. 
These Figures focus on the signal in the LV scenario (indicated by black closed circles), with \( M_{Z^{\prime}} = 50 \) GeV, \( \texttt{g}_{DM} = 1.0 \) and \( \texttt{g}_{l} = 0.003 \), alongside SM backgrounds marked with open colored markers.

The $E_{T}^{miss}$ cut results in a reduction of $t\bar{t}$ and WW events. Similarly, the $|p_{T}^{\mu^{+}\mu^{-}} - E_{T}^{miss}|/p_{T}^{\mu^{+}\mu^{-}}$ cut also lowers the occurrences of $t\bar{t}$, $ZZ(4\mu)$, and WW. 
Additionally, the $\Delta\phi_{\mu^{+}\mu^{-},\vec{E_{T}}^{\text{miss}}}$ cut decreases both $t\bar{t}$ and WW events. The \( \text{cos}(\text{Angle}_{3D})\) cut shows a similar effect, again impacting $t\bar{t}$ and WW. Finally, the \( \Delta R(\mu^{+}\mu^{-})\) cut leads to a decrease in $t\bar{t}$, WW, $ZZ(4\mu)$, and $ZZ(2\mu2\nu)$ events. 
For the signal, we see a flat efficiency across all cuts versus $p_{T}^{\mu}$, with the exception of the $E_{T}^{miss}$ cut, which shows a decline at low $p_{T}^{\mu}$ ($p_{T}^{\mu} < 50$ GeV). 
However, implementing this cut is essential to reduce contamination from WW and $t\bar{t}$ events and to eliminate the process \( e^{+}e^{-} \rightarrow \mu^{+}\mu^{-} \).

\begin{figure}[h!]
\centering
\resizebox*{9.2cm}{!}{\includegraphics{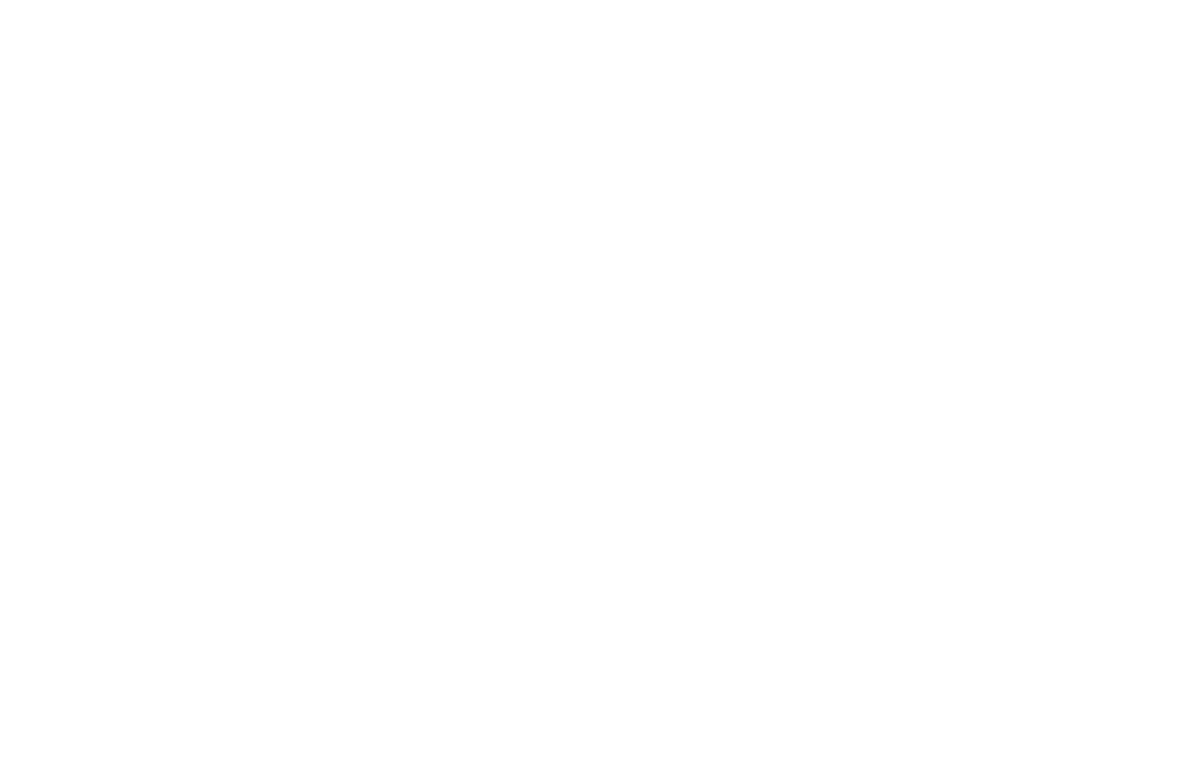}}
\caption{The dimuon invariant mass spectrum for events passing the final selection in Table \ref{cuts}, showing estimated SM backgrounds and various light gauge boson (Z$^{\prime}$) masses from the LV simplified model, with dark matter mass ($M_{\chi_{1}} = 1$ GeV).}
\label{figure:masssemifinal}
\end{figure}

Figure \ref{figure:masssemifinal} displays the dimuon invariant mass spectrum for events that satisfy all the final selection criteria outlined in Table \ref{cuts}, except the cut: \( 0.9 \times M_{Z^{\prime}} < M_{\mu^{+}\mu^{-}} < M_{Z^{\prime}} + 25 \). Additionally, contributions from \(e^+\mu^-\) pairs have been subtracted.
These histograms represent the estimated SM backgrounds and various light neutral gauge boson (Z$^{\prime}$) masses ($M_{Z^{\prime}} =$ 50 and 100 GeV) generated based on the LV simplified model, with a dark matter mass of $M_{\chi_{1}} = 1$ GeV.
\begin{table}
\small
    \centering
    \begin{tabular}{|c|}
\hline
 Final selection \\
\hline
    \hline
  Pre-selection ($\mu$)\\   
  $|p_{T}^{\mu^{+}\mu^{-}}- E_{T}^{\text{miss}}|/p_{T}^{\mu^{+}\mu^{-}} <$ 0.1  \\
   $\Delta\phi_{\mu^{+}\mu^{-},\vec{E}_{T}^{\text{miss}}} >$ 3 rad \\
  
  $\Delta R(\mu^{+}\mu^{-}) < 1.4$ \\
  $\text{cos}(\text{Angle}_{3D})< -0.9$ \\
  $E_{T}^{\text{miss}} > 100$ GeV \\
  $0.9 \times M_{Z^{\prime}} < M_{\mu^{+}\mu^{-}} < M_{Z^{\prime}} + 25$\\
    \hline
    \end{tabular}
    \caption{Summary of cut-based event final selections used in the analysis.}
    \label{cuts}
\end{table}
Based on the final selection analysis previously discussed, the SM background is notably diminished. Additionally, the Z boson background is eliminated. As a result, signal events are distinguishable from the SM background.
\section{Results}
\label{section:Results}
The shape-based analysis employs the distributions of \(\cos\theta_{CS}\) as effective discriminators. These distributions are particularly valuable because the signal characteristic is defined by a typical spin-1 boson pattern, which differs from the hypothesis that a spin-2 graviton decays into a dimuon, as illustrated in \cite{cms-note}. Additionally, the SM background is already negligibly small following the final analysis selection.

The results and all figures displayed in this section, which include \(\cos\theta_{CS}\) and significance plots, have been adjusted by removing the contributions from \(e^+\mu^-\) pairs.

\begin{figure}[h!]
\centering
  \resizebox*{9.2cm}{!}{\includegraphics{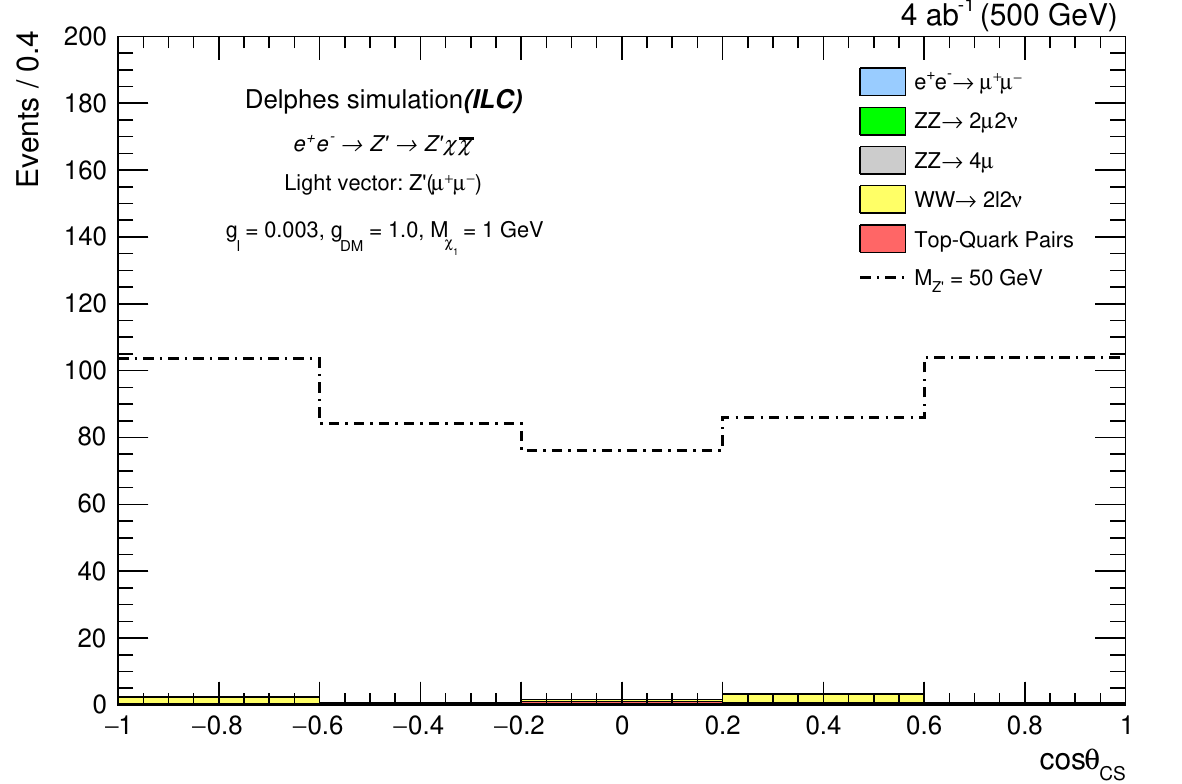}}
  \caption{The distribution of the $\text{cos}\theta_{CS}$, after applying the final analysis listed in Table \ref{cuts}, for the expected SM background and one signal benchmark corresponding to the LV with $M_{Z^{\prime}} = 50$ GeV is superimposed.
  All events are required to have an invariant mass of the dimuon in the range of 45 to 75 GeV.}
  \label{costheta}
\end{figure}

After implementing the final event selection detailed in Table \ref{cuts}, the distribution of $\text{cos}\theta_{CS}$ is presented in Figure \ref{costheta}. 
This plot summarizes the outcomes for both the SM backgrounds and the signal from the LV scenario, corresponding to an integrated luminosity of 4 ab$^{-1}$ at $\sqrt{s} = 500$ GeV. The signal was generated with a light gauge boson mass $M_{Z^{\prime}}$ of 50 GeV and a dark matter mass $M_{\chi_{1}}$ of 1 GeV.

\begin{table*}
\small
    \centering
    
    \begin{tabular}{|c|c|c|}
\hline
Process & No. of events passing pre-selection& No. of events passing final-selection\\
\hline
\hline
$\text{$e^{+}e^{-}$} \rightarrow \mu^{+}\mu^{-}$ & $56734.8 \pm 5678.5$ & $0 \pm 0$\\
$\text{t}\bar{\text{t}} \rightarrow \mu^{+}\mu^{-} + 2b + 2\nu$ & $6292.6 \pm 634.2$ & $1.7 \pm 1.3$\\
$\text{WW} \rightarrow l^{+}l^{-} + 2\nu$ & $9935.3 \pm 998.5$ & $5.0 \pm 2.1$ \\
$\text{ZZ} \rightarrow \mu^{+}\mu^{-} + 2\nu$ & $19.4 \pm 4.8$ & $1.4 \pm 1.2$ \\
$\text{ZZ} \rightarrow 4\mu$ & $245.4 \pm  29.1$ & $0.1 \pm 0.3$ \\
Sum Bkgs $(N_b)$& $73227.5 \pm 7327.7$ & $8.2 \pm 2.8$ \\
\hline
Signal of LV scenario $(N_s)$& $1638.9 \pm 168.8$ & $454.1 \pm 50.2$  \\
(at $M_{Z^{\prime}}$ = 50 GeV and $M_{\chi_{1}} = 1$ GeV) &&  \\
\hline
\hline
Significance & $S_L = 0.4\sigma$ & $S = 21\sigma$ \\
\hline
\end {tabular}
\caption{The Table summarizes the number of events that met the pre-selection (middle column) and full selection (right column) criteria from simulations for SM backgrounds and a signal, conducted with a luminosity of 4 ab$^{-1}$ at $\sqrt{s} = 500$ GeV. The signal corresponds to the LV scenario with $M_{Z^{\prime}}$ = 50 GeV, $M_{\chi_{1}} = 1$ GeV, $g_{DM} = 1.0$, and $g_{l} = 0.003$. The total uncertainties, which include both statistical and systematic components, have been combined using the quadratic form. Signal significance against the SM background is shown before and after the final selection.}
\label{table:tab18}
\end{table*}

Table \ref{table:tab18} outlines the number of events that successfully met both the pre-selection criteria (shown in the middle column) and the full selection criteria (indicated in the right column). These results were derived from simulations that accounted for both backgrounds and a signal, as detailed in the first column. The simulations were performed with a luminosity of 4 ab$^{-1}$ at $\sqrt{s} = 500$ GeV. 
The signal sample is based on the LV scenario, with model parameters set to $M_{Z^{\prime}}$ = 50 GeV, $M_{\chi_{1}} = 1$ GeV, $g_{DM} = 1.0$, and $g_{l} = 0.003$. We incorporated total uncertainties that cover both statistical and systematic components for the simulated signal and background samples. 

When dealing with an enriched sample of events for the SM background and a specific signal process, it is essential to consider the entire distribution of a variable across events, rather than just counting events within a signal region.
In this context, the statistical significance (\( S_L \)) can be defined using the likelihood function, as explained in \cite{SSL, R2}. This method is utilized solely for the SM background and LV single events that meet the pre-selection criteria (refer to the middle column in Table \ref{table:tab18}).

In case of studying a small statistical sample of the SM background, it is important to use the statistical significance defined by the formula \( S = \frac{N_s}{\sqrt{N_s + N_b}} \), in which \( M_{Z^{\prime}} \) is varied. In this equation, \( N_s \) represents the number of signal events from the new physics (LV signal events), while \( N_b \) denotes the total number of SM background events that pass the final selection criteria.

\begin{figure*}
\centering
\hspace{0mm}
\centering
\subfigure[]{
  \includegraphics[width=72.mm]{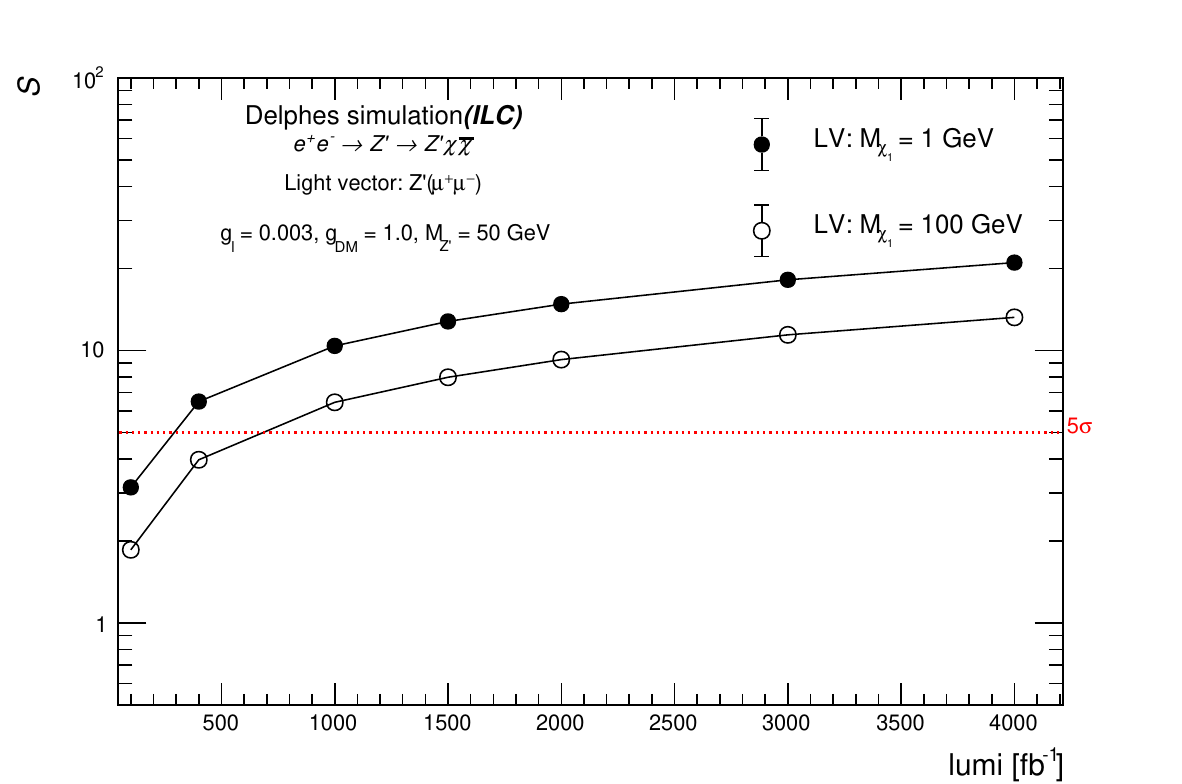}
  \label{figure:s2}
}
\hspace{0mm}
\centering
\subfigure[]{
  \includegraphics[width=72.mm]{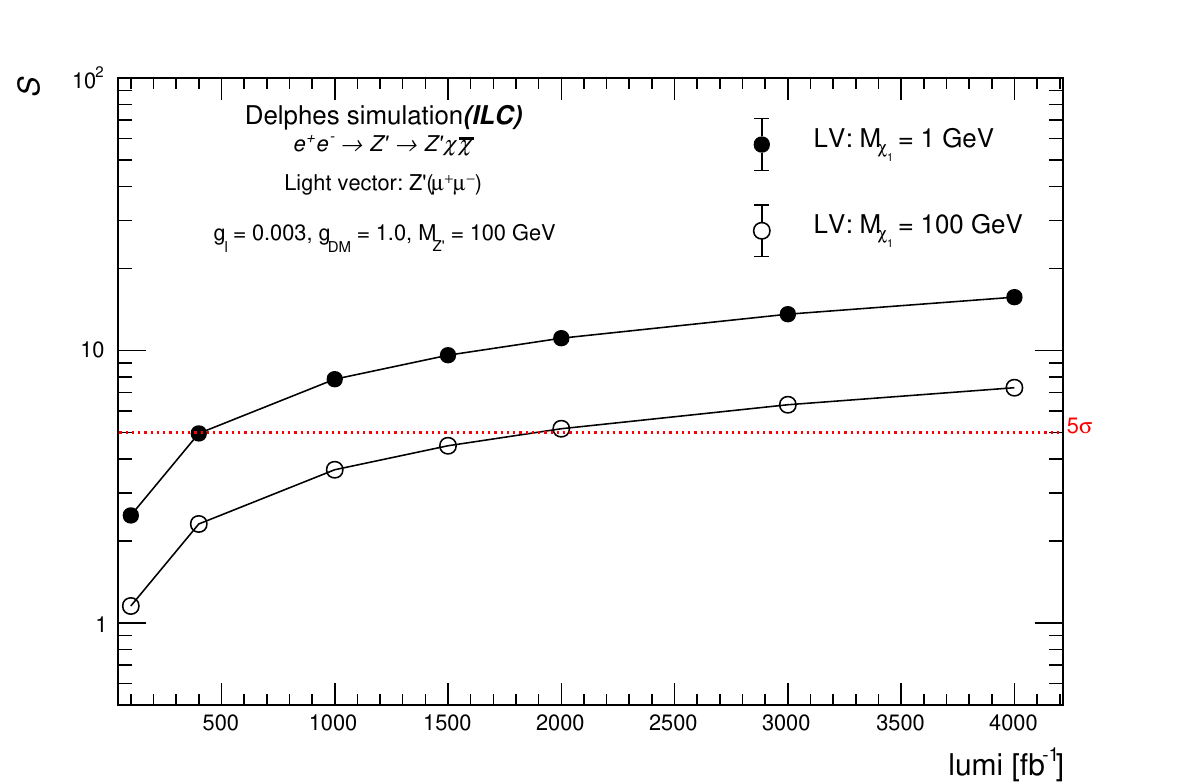}
  \label{figure:s3}
}
\caption{The statistical significance ($\it{S}$) versus integrated luminosity for $M_{Z^{\prime}} = 50$ GeV (Figure \ref{figure:s2}) and 100 GeV (Figure \ref{figure:s3}) is shown with varying dark matter masses ($M_{\chi_{1}}$) for events passing the final cuts in Table \ref{cuts}. The model signal represents the LV scenario with coupling constants $\texttt{g}_{l} = 0.003$, and $\texttt{g}_{DM} = 1.0$ at $\sqrt{s} = 500$ GeV. The dashed red line indicates $\it{S} = 5$.}
\label{figure:significance2}
\end{figure*}

Figures \ref{figure:s2} and \ref{figure:s3} illustrate the statistical significance (\(S\)) as a function of integrated luminosity for two scenarios of dark matter mass (\(M_{\chi_{1}} = 1\) GeV and \(100\) GeV). Figure \ref{figure:s2} focuses on \(M_{Z^{\prime}} = 50\) GeV, while Figure \ref{figure:s3} focuses on \(M_{Z^{\prime}} = 100\) GeV, for events that meet final criteria outlined in Table \ref{cuts}.
These figures reflect the model associated with the LV scenario, using coupling constants of \(\texttt{g}_{l} = 0.003\) and \(\texttt{g}_{DM} = 1.0\) at $\sqrt{s} = 500$ GeV. The dashed red line in the plots represents a significance value of \(S = 5 \).

For a mass \( M_{Z^{\prime}} \) of 50 GeV as in \ref{figure:s2}, one can achieve a 5$\sigma$ discovery at an integrated luminosity of 293 fb\(^{-1}\) for low-mass dark matter (\( M_{\chi_1} = 1 \) GeV). In comparison, obtaining this for heavy dark matter (\( M_{\chi_1} = 100 \) GeV) requires an integrated luminosity of 688 fb\(^{-1}\).

For a \( M_{Z^{\prime}} \) of 100 GeV, as shown in \ref{figure:s3}, a significant of 5$\sigma$ discovery can be achieved with an integrated luminosity of 400 fb\(^{-1}\) when considering low-mass dark matter (\( M_{\chi_1} = 1 \) GeV). On the other hand, if we look at heavier dark matter, specifically with a mass of \( M_{\chi_1} = 100 \) GeV, a 5$\sigma$ discovery would require reaching an integrated luminosity of 1893 fb\(^{-1}\).

We employed the profile likelihood method to statistically analyze our results, focusing specifically on the distributions of $\text{cos}\theta_{CS}$. To establish exclusion limits on the product of signal cross sections and the branching fraction Br($Z^{\prime}$ $\rightarrow \mu\mu$) at a 95\% confidence level, we utilized the modified frequentist CLs construction \cite{R58, R59}, which relies on the asymptotic approximation \cite{R2}. 
A likelihood ratio is used as the test statistic, and systematic uncertainties are treated as nuisance parameters.

\begin{figure}[htb]
\centering
  \resizebox*{10.0cm}{!}{\includegraphics{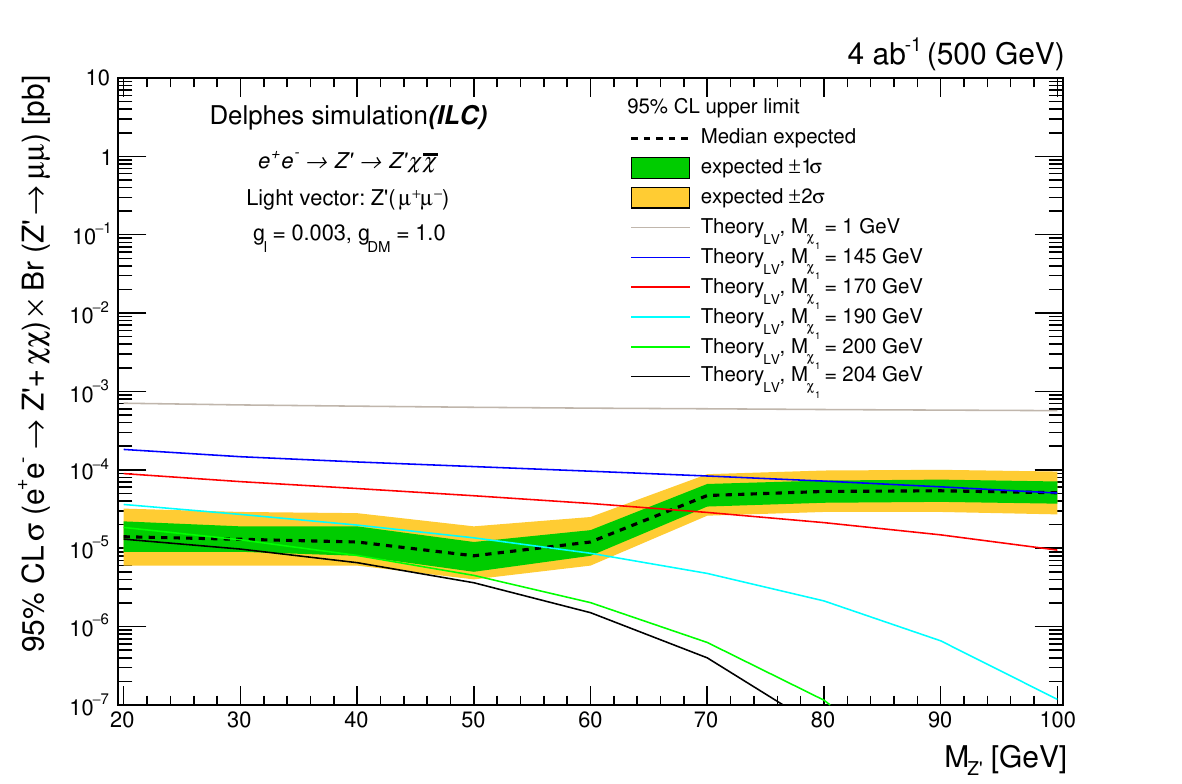}}
\caption{The expected 95\% CL upper limits on the cross-section times branching ratio as a function of the $Z'$ mass ($M_{Z^{\prime}}$) for the LV scenario based on the mono-Z$^{\prime}$ model with its muonic decay. Solid colored lines correspond to the LV scenario with taking $M_{\chi_{1}} =$ 1, 145, 170, 190, 200, and 204 GeV.}
\label{figure:fig7}
\end{figure}

\begin{figure}[htb]
\centering
  \resizebox*{10.0cm}{!}{\includegraphics{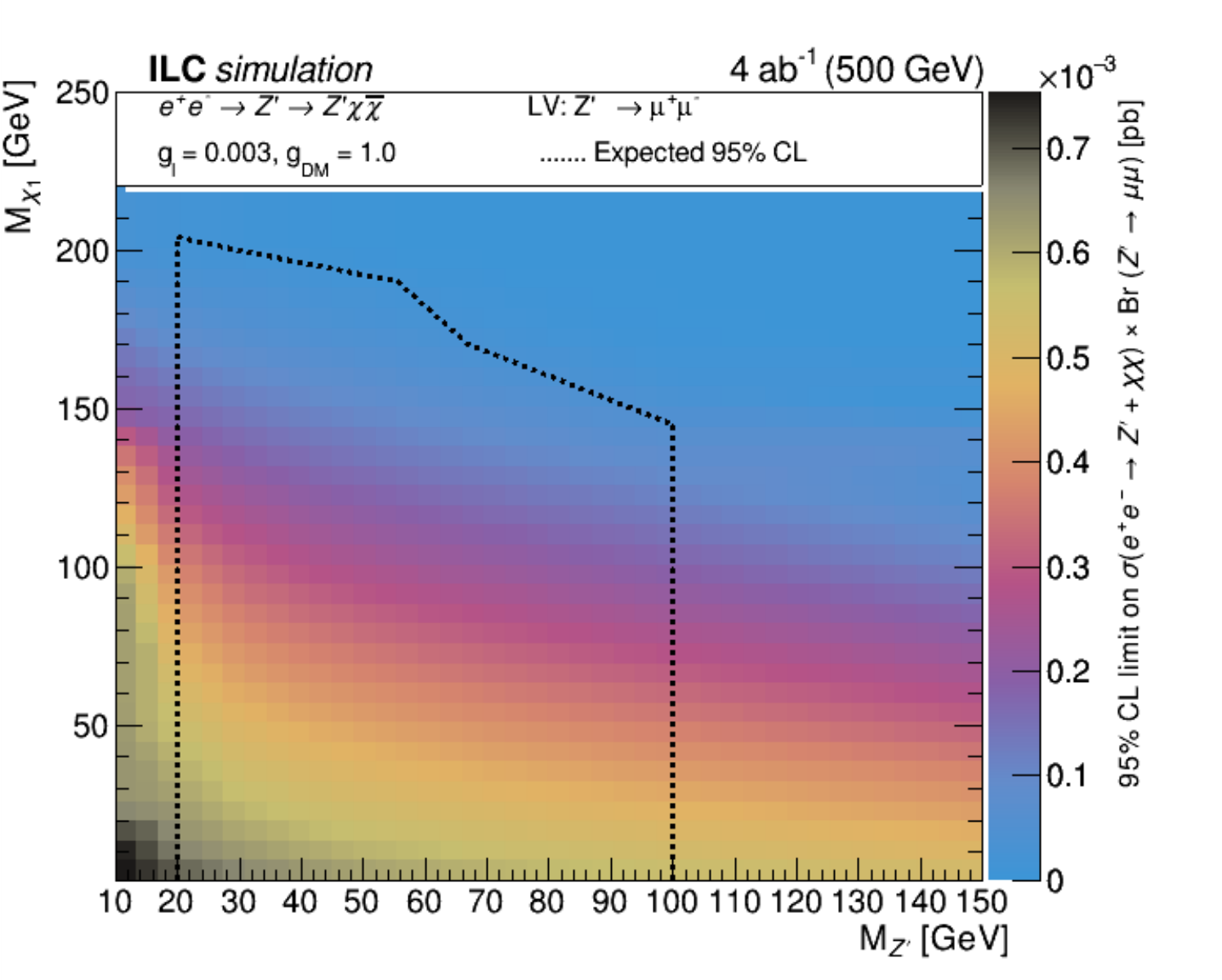}}
  \caption{The 95\% CL upper limits on the cross-section times branching ratio from the search for varying pairs of LV scenario parameters ($M_{Z^{\prime}}$ and $M_{\chi_{1}}$) are shown. The filled region indicates the upper limit, and the dotted black curve marks the expected exclusions for the nominal Z$^{\prime}$ cross-section with coupling constants $\texttt{g}_{l} = 0.003$, and $\texttt{g}_{DM} = 1.0$ at $\sqrt{s} = 500$ GeV.}
  \label{figure:fig8}
\end{figure}

Figure \ref{figure:fig7} presents the anticipated 95\% upper limit on the product of the cross-section and the branching ratio for the LV scenario, specifically focusing on the muonic decay of the Z$^{\prime}$. This analysis is based on coupling constant values of $\texttt{g}_{l} = 0.003$ and $\texttt{g}_{DM} = 1.0$, derived from an integrated luminosity of 4 ab$^{-1}$ at a center-of-mass energy of $\sqrt{s} = 500$ GeV. The limits are depicted by distinct colored solid lines corresponding to various dark matter mass values of \(M_{\chi_{1}}\), including 1, 145, 170, 190, 200, and 204 GeV. In contrast, Figure \ref{figure:fig8} shows the limits on the product of cross-sections and branching ratios for the muonic decay channel of the Z$^{\prime}$ boson, plotted as a function of both the mediator's mass ($M_{Z^{\prime}}$) and the mass of the dark matter particle ($M_{\chi_{1}}$). The area bounded by the contour indicates the regions that are excluded at the 95\% confidence level for the benchmark scenario with $\texttt{g}_{l} = 0.003$ and $\texttt{g}_{DM} = 1.0$. 

This expected limit shows that Z$^{\prime}$ mass ranging from 20 to 100 GeV can be excluded for $M_{\chi_{1}} \in [1, 145]$ GeV for an integrated luminosity of 4 ab$^{-1}$ at $\sqrt{s} =500$ GeV.
For high dark matter mass values, specifically when \( M_{\chi_1} > 204 \) GeV, the ILC will not be sensitive to the LV scenario characterized by \( g_l = 0.003 \) and \( g_{DM} = 1.0 \). 


\section{Summary}
\label{section:Summary}
The ILC is a foreseen electron-positron collider that will be designed for discovering particles that go beyond the standard model (BSM). It offers a distinct signature that helps identify unknown particles, such as dark matter, additional neutral gauge bosons (\( Z^{\prime} \)), and Kaluza-Klein excitations, even when faced with the background noise of quantum chromodynamics (QCD).

Our research investigated the angular distributions of low-mass dimuon pairs within the Collins-Soper frame, using simulated data samples from the ILC. These Monte Carlo (MC) samples were generated from electron-positron collisions with a center-of-mass energy of 500 GeV, including signal and standard model background events. This setup corresponds to what is anticipated for ILC Run 1, which is expected to have an integrated luminosity of 4 ab\(^{-1}\).
Our analysis focused on the cos\(\theta_{\text{CS}}\) variable to extract valuable insights from the MC data. 
This variable highlights the spin characteristics of the \( Z^{\prime} \) spin-1 model, which is essential for distinguishing it from other models, such as the spin-2 graviton, by utilizing this distribution.

In this study, we investigated the effects of a simplified model scenario known as the light vector, focusing on dark matter pair production associated with a low mass \( Z^{\prime} \) boson ($M_{Z^{\prime}} < 100$ GeV) at the ILC, something the LHC could not access. We considered the muonic decay of the \( Z^{\prime} \) boson, with coupling constants fixed at \( g_{DM} = 1.0 \) and \( g_{l} = 0.003 \). 

We implemented effective discrimination cuts that successfully eliminated the Z boson background, allowing us to better differentiate between signal events and SM backgrounds. As a result, we observed a significant reduction in SM backgrounds while preserving the signal strength by applying appropriate cuts, which are detailed in Table \ref{cuts} for the light vector scenario.

For \(M_{Z^{\prime}} = 50\) GeV, by employing these strong selection criteria, a substantial integrated luminosity of 293 fb\(^{-1}\) allows for the potential discovery of low-mass dark matter, specifically with a mass of \(M_{\chi_1} = 1\) GeV, achieving a significance of 5$\sigma$. In contrast, to observe heavier dark matter with a mass of \(M_{\chi_1} = 100\) GeV, a significantly larger integrated luminosity of 688 fb\(^{-1}\) would be necessary.

When \(M_{Z^{\prime}} = 100\) GeV, a 5$\sigma$ discovery can be reached with an integrated luminosity of 400 fb\(^{-1}\), after the first four years of ILC running at \(\sqrt{s} = 500\) GeV, for low-mass dark matter (\(M_{\chi_1} = 1\) GeV). However, for heavy dark matter (\(M_{\chi_1} = 100\) GeV), a 5$\sigma$ discovery is attainable with an integrated luminosity of 1.9 ab\(^{-1}\).
 
In case this precise signal is not detected at the ILC, we have established upper limits on the masses of \( Z' \) and dark matter (\( \chi_1 \)) at the 95\% confidence level for the muonic decay of \( Z' \).
Thus, for the LV scenario with \( g_l = 0.003 \) and \( g_{DM} = 1.0 \), we have set expected limits that eliminate $M_{Z'}$ range from 20 to 100 GeV for \( M_{\chi_1} \) within the range of 1 to 145 GeV. Notably, it also excludes \( M_{\chi_1} = 204 \) GeV when \( M_{Z'} \) is at 20 GeV.

The prospect of uncovering dark matter through future electron-positron colliders, particularly at the International Linear Collider (ILC), brings a specific focus on mono-photon analyses and the Light vector (LV) scenario in the framework of mono-\( Z' \). 
At a collision energy of 500 GeV, the ILC has the potential to detect Weakly Interacting Massive Particles (WIMPs) with masses up to 250 GeV, achieving a 95\% level of confidence via mono-photon analysis. However, in the LV scenario, where particular coupling constants are set \( g_l = 0.003 \) and \( g_{DM} = 1.0 \), detection capabilities are limited to particles weighing no more than 204 GeV. In contrast, previous experiments such as LEP-2 were only able to exclude much lighter dark matter candidates, establishing a lower mass limit of around 40 GeV based on certain theoretical assumptions.
\\
\\

\begin{acknowledgments}
The author of this paper expresses gratitude to Tongyan Lin, co-author of \cite{R1}, for providing the UFO model files, assisting in the generation of signal events, and cross-checking the results. 
Additionally, this research was supported by the Science, Technology, and Innovation Funding Authority (STDF) under grant number 48289.
\end{acknowledgments}

\textbf{Data Availability Statement:} This manuscript has no associated data or the data will not be deposited. 


\end{document}